\documentclass[
reprint,
 amsmath,amssymb,
 aps,
floatfix,
]{revtex4-2}

\usepackage{graphicx}% Include figure files
\usepackage{dcolumn}% Align table columns on decimal point
\usepackage{bm}% bold math
\usepackage{url}
\usepackage{array}
\usepackage{multirow}% Multiple rows in table
\newenvironment{conditions}
  {\par\vspace{\abovedisplayskip}\noindent\begin{tabular}{>{$}l<{$} @{${}={}$} l}}
  {\end{tabular}\par\vspace{\belowdisplayskip}}

\begin{document}

%%%%%%%%%%%%%%%%%%%%%%%%%%%%%%%%%%%%%%%%%%%%%%%%%%%%%%%%%%%%%%%%%%%%%%
%%%%%%%%%%%%%%%%%%%%%%%%% Title and Abstract %%%%%%%%%%%%%%%%%%%%%%%%%
%%%%%%%%%%%%%%%%%%%%%%%%%%%%%%%%%%%%%%%%%%%%%%%%%%%%%%%%%%%%%%%%%%%%%%

\title{A systematic study of projection biases in the Weak Lensing analysis of cosmic shear and the combination of galaxy clustering and galaxy-galaxy lensing.}

\author{P.R.V.~Chintalapati}
\email{raj.ch90@gmail.com}
 \affiliation{Northern Illinois University}
 \affiliation{Fermi National Accelerator Laboratory, P.O. Box 500, Batavia, IL 60510, USA.}
\author{G.~Gutierrez}
 \email{gaston@fnal.gov}
\affiliation{Fermi National Accelerator Laboratory, P.O. Box 500, Batavia, IL 60510, USA.}
\author{M.H.L.S.~Wang}
 \email{mwang@fnal.gov}
\affiliation{Fermi National Accelerator Laboratory, P.O. Box 500, Batavia, IL 60510, USA.}

\date{\today}

\begin{abstract}
This paper presents the results of a systematic study of projection biases in the Weak Lensing analysis of cosmic shear and the combination of galaxy clustering and galaxy-galaxy lensing using data collected during the first-year of running the Dark Energy Survey experiment.  The study uses $\Lambda$CDM as the cosmological model and two-point correlation functions for the WL analysis. The results in this paper show that, independent of the WL analysis, projection biases of more than $1\sigma$ exist, and are a function of the position of the true values of the parameters $h$, $n_{s}$, $\Omega_{b}$, and $\Omega_{\nu}h^{2}$ with respect to their prior probabilities. For cosmic shear, and the combination of galaxy clustering and galaxy-galaxy lensing, this study shows that the coverage probability of the $68.27\%$ credible intervals ranges from as high as $93\%$ to as low as $16\%$, and that these credible intervals are inflated, on average, by $29\%$ for cosmic shear and $20\%$ for the combination of galaxy clustering and galaxy-galaxy lensing. The results of the study also show that, in six out of nine tested cases, the reduction in error bars obtained by transforming credible intervals into confidence intervals is equivalent to an increase in the amount of data by a factor of three.
\end{abstract}

\maketitle

%\tableofcontents

%%%%%%%%%%%%%%%%%%%%%%%%%%%%%%%%%%%%%%%%%%%%%%%%%%%%%%%%%%%%%%%%%%%%%%
%%%%%%%%%%%%%%%%%%%%%%%%%%%% Introduction %%%%%%%%%%%%%%%%%%%%%%%%%%%%
%%%%%%%%%%%%%%%%%%%%%%%%%%%%%%%%%%%%%%%%%%%%%%%%%%%%%%%%%%%%%%%%%%%%%%

\section{Introduction}\label{sec:introduction}

Weak Lensing (WL) analysis, which uses galaxy positions and shapes to measure cosmological parameters, is a powerful tool available today to probe the structure of the universe and its evolution.  
Since its first significant measurement in 2000 \cite{Sloan_WL,Nature_2000_WL,MNRAS_2000_WL,A&A_2000_WL},WL analyses have evolved to utilize the combination of three two-point correlation functions: (i) the correlation of galaxy number densities along two different lines of sight (LOS), referred to as galaxy clustering, (ii) the correlation of galaxy number densities along a LOS with the galaxy shear along a different LOS, known as galaxy-galaxy lensing, and (iii) the correlation of galaxy shear along two different LOS, or cosmic shear. Sky surveys like the Dark Energy Survey (DES) \cite{DES_WEB} use WL analyses to measure cosmological parameters like the total relative matter density $\Omega_m$, the normalization of the power spectrum $\sigma_8$, and their combination $S_8 = \sigma_8 \sqrt{\Omega_m / 0.3}$. 

The combined analysis using all three two-point correlation functions is referred to as 3x2pt analysis, and provides the best constraining power for the cosmological parameters extracted from weak lensing. However, as shown by DES recently \cite{DES_Y3_1x2pt-a,DES_Y3_1x2pt-b,DES_Y3_2x2pt-a,DES_Y3_2x2pt-b}, significant constraining power can also be obtained using only cosmic shear, known as a 1x2pt analysis, and the combined analysis of galaxy clustering and galaxy-galaxy lensing, or a 2x2pt analysis.  It is, therefore, crucial to have a detailed understanding of the systematic and statistical biases for each type of WL analysis: 3x2pt, 2x2pt, and 1x2pt. This is especially important as we enter the era of “Stage IV” sky surveys, like Euclid \cite{Euclid_WEB}, LSST \cite{LSST_WEB}, and the Nancy Roman Space Telescope \cite{NRST_WEB}, which will significantly increase the constraints on cosmological parameters in comparison to current surveys.  The study of projection biases in the 3x2pt analysis has already been published\cite{PRD_us}.  In this paper, we will focus our attention on the projection biases in the 2x2pt and 1x2pt WL analysis.

All current WL analyses, whether based on 1x2pt, 2x2pt, or 3x2pt correlation functions, use Bayesian statistics.  The Bayesian statistical method is used to determine posterior probabilities (or posteriors) of the parameters to be measured. The interval that covers a fraction of the posterior probability (e.g. a $68$\%) is specified as a credible interval (e.g. a $68$\% credible interval). The probability value of this credible interval represents the plausibility or degree of belief in finding the true value of a parameter within that interval. These credible intervals differ conceptually from the confidence intervals used in the frequentist method. In the frequentist method, a $68 \,(95)$\% confidence interval is constructed such that, when the experiment is repeated many times, the interval contains the true value of the parameter $68 \,(95)$\% of the time. This property is referred to as the ``coverage probability," or simply ``coverage," of the confidence interval. The measured parameter values in Bayesian analysis are obtained using marginalized posteriors with their peaks as the most likely values and the widths of the credible intervals as the errors. Marginalized parameter posteriors are affected by the choice of prior probabilities. Therefore, it is important to study and understand how the priors affect the most likely values obtained from Bayesian analysis. By construction, credible intervals are not guaranteed to have the same coverage as confidence intervals. It is therefore also useful to study the coverage of credible intervals and understand how to transform credible intervals into confidence intervals.

A previous study \cite{PRD_us} of the WL 3x2pt analysis of the DES year one (Y1) data  quantitatively showed that there are projection biases greater than $1\sigma$ in the peaks of the Bayesian posterior probabilities for $\Omega_m$, and greater than $1.5\sigma$ for $\sigma_8$ and $S_8$, due to the poorly constrained and correlated parameters $h$, $n_{s}$, $\Omega_{b}$, and $\Omega_{\nu}h^{2}$. The study showed that these biases depend on the position of the true values of $h$, $n_{s}$, $\Omega_{b}$, and $\Omega_{\nu}h^{2}$ with respect to their prior interval, and that the coverage probabilities of the 68.27\% credible intervals of the parameters $\Omega_m$, $\sigma_8$ and $S_8$ could be as low as 42.27\%, 26.36\%, and 31.36\% respectively.

For the DES Y1 1x2pt and 2x2pt WL analysis, this paper will, 1) quantitatively describe the biases in the posterior peaks and credible intervals of $\Omega_m$, $\sigma_8$, and $S_8$, 2) show that transforming the credible intervals of the previous parameters into confidence intervals can produce gains equivalent to increasing the amount of data by almost three times, and 3) show how the priors distort the usual $1/\sqrt{n}$ scaling of errors when the amount of data is increased by a factor of $n$.

Section \ref{sec:Y1_analysis} briefly discusses the 1x2pt and 2x2pt analysis used in this study and presents the details of the code used for the study. Sections \ref{sec:ProjectionBias} \& \ref{subsec:quantifying} discuss how the biases behave for a simple case and the methods we use to calculate biases. Sections \ref{subsec:peaks} \& \ref{subsec:intervals} present the results of the projection bias study in the peaks of the posteriors and credible intervals for the 2x2pt and the 1x2pt analysis. Section \ref{sec:behavior} explains how the credible intervals scale with an increase in statistics. Section \ref{sec:LargerStatistics} estimates the biases for larger data sets. 

%%%%%%%%%%%%%%%%%%%%%%%%%%%%%%%%%%%%%%%%%%%%%%%%%%%%%%%%%%%%%%%%%%%%%%
%%%%%%%%%%%%%% Describe the Y1 WL package that we used  %%%%%%%%%%%%%%
%%%%%%%%%%%%%%%%%%%%%%%%%%%%%%%%%%%%%%%%%%%%%%%%%%%%%%%%%%%%%%%%%%%%%%

\section{The DES Y1 WL Analysis}\label{sec:Y1_analysis}

For the studies presented in this paper, we will use the same code used to analyze the WL data collected during the first year of observation by DES \cite{DES_Y1_PRD}, and also used for the projection bias study for the DES Y1 3x2pt WL analysis \cite{PRD_us}.  This code and the data are publicly available in CosmoSIS \cite{Cosmosis_paper,Cosmosis_program}.  Furthermore, the projection bias studies presented here are performed using a $\Lambda$CDM fiducial cosmology, with the cosmological parameters extracted using two different analyses. As mentioned earlier in the 2x2pt analysis, the parameters are extracted by fitting two two-point correlation functions, Galaxy clustering and Galaxy-galaxy lensing. In the 1x2pt analysis, the parameters are extracted by fitting to one two-point correlation function, Cosmic shear. In both 1x2pt and 2x2pt analysis, the two-point correlations are given as a function of the angle $\theta$ separating the two LOS, and in bins of redshift along the LOS.  In Subsection \ref{subsec:Y1_model}, we will give a short description of how the two-point correlation functions are calculated using the power spectrum and the galaxy number densities. In Subsection \ref{subsec:Y1_code}, we will describe the code that was used in the analysis and how we selected the ``true" or nominal values of the parameters for our studies.

\subsection{The Model}\label{subsec:Y1_model}

The nonlinear power spectrum $\hat{P}_{NL} (k,z)$, where $k$ is the wave number and $z$ is the redshift, was calculated using the CosmoSIS pipeline (which includes CAMB\cite{CAMB} \& Halofit\cite{Halofit_1,Halofit_2,Halofit_3}), and the seven cosmological parameters listed in Table \ref{tab:parameters}.  DES uses a flat sky and the Limber approximations \cite{DES_Y1_PRD} for their Y1 data analysis.  In the Limber approximation only modes perpendicular to the line of sight enter in the calculation of the power spectrum, that is $\hat{P}_{NL} (k_{\perp},z)$, with $k_\perp = (l+1/2)/\chi$, and where $l$ is a positive number and $\chi$ is the comoving distance along the LOS.  The two-point correlation functions are measured separately for each bin of redshift and therefore, to calculate these correlations, galaxy number densities are needed for each redshift bin. DES uses two sets of galaxy samples: lens galaxies that are divided into five redshift bins and source galaxies that are divided into four redshift bins. The lens galaxies are used to calculate clustering while the source galaxies also have information about their shape, which is used to measure shear for those galaxies. Including measurement uncertainties, and as a function of redshift bins, the normalized number densities for the lenses $\hat{n}_{g}^i(z)$, and the sources $\hat{n}_{k}^i(z)$, are given by

\begin{equation}
\hat{n}_{g/\kappa}^i(z) = \hat{n}_{PZ,g/\kappa}^i(z_{l/s}-\Delta z_{l/s}^i)
\end{equation}
where $\hat{n}_{PZ,g/\kappa}^i$ is the measured number density in $i^{th}$ redshift bin, and $\Delta z_{l}^i$ and $\Delta z_{s}^i$ are the lens and source nuisance parameters, respectively, used to account for the measurement errors in redshift.  These errors are varied during the analysis, subject to the priors listed in Tables \ref{tab:parameters} and \ref{tab:parameters2}. For the calculations in this study, the relation between redshift and comoving distance $\chi$, $z = z(\chi)$, is calculated using a $\Lambda$CDM model, and the number density in the $i^{th}$ bin as a function of comoving distance is given by

\begin{equation}
n^i_{g/\kappa}(\chi) = \hat{n}^i_{g/\kappa}(z) \; \frac{d z}{d \chi}
\end{equation}
In the following sections, we will briefly discuss how, using the nonlinear power spectrum and the galaxy number densities, the two-point correlation functions are calculated for the WL analysis of DES Year 1 data.

\subsubsection{Cosmic Shear ($1\times2$pt) analysis}\label{subsubsec:cosmic_shear}

Cosmic shear, as defined in the introduction, is given as a function of the angular separation $\theta$ between the two Lines of Sight (LOS). The tangential and cross shear two-point correlations are calculated separately, and then the sum and difference of the two are used to form two new quantities $\xi_{+/-}(\theta)$ which are theoretically calculated as

\begin{align}
\xi_{+/-}^{i j}(\theta) &= (1+m^i) \, (1+m^j) \, \widehat{\xi}_{+/-}^{\phantom{i}i j}(\theta)\\
\widehat{\xi}_{+/-}^{\phantom{i}i j}(\theta) &= \!\! \int \frac{dl}{2\pi} \; l J_{0/4}(l \theta) \! \int \frac{d\chi}{\chi^2} \, q^i(\chi) \, q^j(\chi) \,  \hat{P}_{NL}(k_\perp,z)
\label{eq:shear-shear}
\end{align}
where the indices $i$ and $j$ label the redshift bins of the lenses and the sources, respectively, and
\begin{conditions}
 m^{i}     &  $i^{th}$ bin error in galaxy shear measurements \\
 q^i(\chi) &  $\widehat{q}^{\; i}(\chi)$ - IA correction \\
 \widehat{q}^{\; i}(\chi) & Lensing efficiency in the $i^{th}$ bin \\
 J_{n}(x) & Bessel functions
\end{conditions}
The nuisance parameters $m^{i}$ account for measurement errors in the shear measurements.  These parameters are allowed to float during the analysis constrained by their priors.  The lensing efficiency is calculated using the galaxy number density as
\begin{equation}
\widehat{q}^{\phantom{i}i}(\chi) = \frac{3 \, H_0^2 \, \Omega_m}{2 c^2 \, a(\chi)} \int_\chi^{\chi_h} d\chi^s \, n^i_\kappa(\chi^s) \; \chi \left( 1 - \frac{\chi}{\chi^s} \right)
\label{eq:lensing_efficiency}
\end{equation}
where $H_0$ is the Hubble constant, $c$ is the speed of light, $a$ is the scale factor, and $\chi_h$ is the comoving distance at the horizon. In the DES Y1 analysis, the lensing efficiency was modified to account for the intrinsic alignment (IA) of galaxies \cite{Hirata-Seljak, Hirata-Seljak-erratum}.  The ``non-linear linear alignment" (NLA) model \cite{Bridle-King_IA} was used to calculate this correction, which is given by
\begin{equation}
\text{IA correction} = A_{IA} \left( \frac{1+z}{1+z_0} \right)^{\eta_{IA}} \frac{0.0134 \, \Omega_m}{D(\chi)} \; n^i_\kappa(\chi)
\label{eq:IA_corr}
\end{equation}
where $z_0=0.62$ and $D(\chi)$ is the linear growth factor,  and the intrinsic alignment parameters $A_{IA}$ and $\eta_{IA}$ enter the analysis as part of the nuisance parameters. The nominal values and priors of all 16 parameters used in the 1x2pt analysis are listed in Table \ref{tab:parameters}.
\begin{table}[hbt]
\caption{\label{tab:parameters}%
This Table lists the nominal values, ranges, and priors of the parameters used in the projection bias studies of DES Y1 1x2pt WL analysis \cite{DES_Y1_PRD}. Parameters with the nominal values labeled as ``varied" indicate that,  for the projection bias studies in this paper, they assume the different combinations of values listed in Table \ref{tab:variables}. The shape of the prior, within the specified range, is given as either ``Flat" or ``Gauss($\mu,\sigma$)" representing either a flat or a Gaussian prior with mean $\mu$ and standard deviation $\sigma$.}
\begin{ruledtabular}
\begin{tabular}{lccc}
\textrm{Parameter}&
\textrm{Nominal} & \multicolumn{1}{c}{\textrm{Range}}&
\textrm{Prior}\\
\colrule
\multicolumn{4}{c}{\textrm{Cosmological Parameters}} \\
$\Omega_m$ & 0.267 & (0.10, 0.90) & Flat \\
$A_s \times 10^9$ & 2.870 & (0.50, 5.00) & Flat \\
$n_s$ & varied & (0.87, 1.07) & Flat \\
$\Omega_b$ & varied & (0.03, 0.07) & Flat \\
$h$ & varied & (0.55, 0.91) & Flat \\
$\Omega_\nu h^2 \times 10^3$ & varied & (0.50, 10.0) & Flat \\
$w$ & -1 & fixed & \\
\hline
\multicolumn{4}{c}{\textrm{Intrinsic Alignment Parameters}} \\
$A_{IA}$ & 0.44 & (-5.0,5.0) & Flat \\
$\eta_{IA}$ & -0.7 & (-5.0,5.0) & Flat \\
\hline
\multicolumn{4}{c}{\textrm{Uncertanities in source redshifts}} \\
$\Delta z_s^1 \times 10^2$ & -0.1 & (-10.,10.) & Gauss(-0.1,1.6) \\
$\Delta z_s^2 \times 10^2$ & -1.9 & (-10.,10.) & Gauss(-1.9,1.3) \\
$\Delta z_s^3 \times 10^2$ & \phantom{-}0.9 & (-10.,10.) & Gauss(\phantom{-}0.9,1.1) \\
$\Delta z_s^4 \times 10^2$ & -1.8 & (-10.,10.) & Gauss(-1.8,2.2) \\
\hline
\multicolumn{4}{c}{\textrm{Shear bias parameters}} \\
$m^{i (=1,4)} \times 10^2 $ & \phantom{-}1.2 & (-10.,10.) & Gauss(\phantom{-}1.2,2.3) \\
\end{tabular}
\end{ruledtabular}
\end{table}
\subsubsection{Galaxy clustering and Galaxy-galaxy lensing ($2\times2$pt) analysis}\label{subsubsec:clustering}

In the combined analysis of galaxy clustering and galaxy-galaxy lensing, the auto-correlation function $w(\theta)$ of galaxy number densities, and the correlation function $\gamma_t(\theta)$ of galaxy number densities with galaxy shears, are theoretically calculated as follows.
The galaxy clustering is calculated correlating the galaxy number densities in bins of redshift along two different lines of sight separated by an angle $\theta$, the result is given by
\begin{equation}
\widehat{w}^{\phantom{l}i}(\theta) = \int \frac{dl}{2\pi} \; l J_0(l \theta) \int \frac{d\chi}{\chi^2} \; [n^i_g(\chi)]^2 \, \hat{P}_{NL}(k_\perp,z)
\label{eq:gal-clus}
\end{equation}
Since $n^i_g(\chi)$ only accounts for visible matter, a galaxy bias factor $b_i$ needs to be included for each lens redshift bin to account for non-visible matter. Therefore, the galaxy clustering two-point correlation function is 
\begin{equation}
w^i(\theta) = b_i^2 \, \widehat{w}^{\phantom{l}i}(\theta)
\label{eq:gal-clus-final}
\end{equation}
The galaxy bias factors are allowed to float during the analysis while constrained by the priors given in Table \ref{tab:parameters2}.  Galaxy-galaxy lensing is obtained by calculating the correlations between the galaxy number densities in lens redshift bins and the shear measurement of the galaxies in source redshift bins. This correlation is theoretically calculated as 
\begin{equation}
\widehat{\gamma}_t^{\phantom{i}i j}(\theta) = \int \frac{dl}{2\pi} \; l J_2(l \theta) \int \frac{d\chi}{\chi^2} \; n^i_g(\chi) \, q^j(\chi) \, \hat{P}_{NL}(k_\perp,z)
\label{eq:gal-shear}
\end{equation}
The final Galaxy-galaxy lensing correlation includes the galaxy bias $b_i$ and the shear measurement bias $m^j$, which leads to
\begin{equation}
\gamma_t^{i j}(\theta) = b_i \, (1+m^j) \, \widehat{\gamma}_t^{\phantom{i}i j}(\theta)
\label{eq:gal-shear-final}
\end{equation}
In addition to the $16$ parameters used in the DES Y1 $1\times2$pt analysis, there are $10$ more parameters that are used in the $2\times2$pt analysis. 
These $10$ parameters include the galaxy bias parameters and the uncertainties in the lens redshift distributions whose nominal values and priors are listed in Table \ref{tab:parameters2}.

\begin{table}[hbt]
\caption{\label{tab:parameters2}%
Table showing the additional parameters to the ones listed in Table \ref{tab:parameters}, that were used in the DES Y1 2x2pt WL analysis \cite{DES_Y1_PRD}, together with their nominal values, range, and priors.  ``Flat" denotes a prior that is flat within the given range and ``Gauss($\mu,\sigma$)" specifies a Gaussian prior with mean $\mu$ and standard deviation $\sigma$.}
\begin{ruledtabular}
\begin{tabular}{lccc}
\textrm{Parameter}&
\textrm{Nominal} & \multicolumn{1}{c}{\textrm{Range}}&
\textrm{Prior}\\
\colrule
\multicolumn{4}{c}{\textrm{Galaxy bias parameters}} \\
$b_1$ & 1.42 & (0.8, 3.) & Flat \\
$b_2$ & 1.65 & (0.8, 3.) & Flat \\
$b_3$ & 1.60 & (0.8, 3.) & Flat \\
$b_4$ & 1.92 & (0.8, 3.) & Flat \\
$b_5$ & 2.00 & (0.8, 3.) & Flat \\
\hline
\multicolumn{4}{c}{\textrm{Uncertanities in lens redshifts}} \\
$\Delta z_l^1 \times 10^2$ & \phantom{-}0.8 & (-5.0,5.0) & Gauss(\phantom{-}0.8,0.7) \\
$\Delta z_l^2 \times 10^2$ & -0.5 & (-5.0,5.0) & Gauss(-0.5,0.7) \\
$\Delta z_l^3 \times 10^2$ & \phantom{-}0.6 & (-5.0,5.0) & Gauss(\phantom{-}0.6,0.6) \\
$\Delta z_l^4 \times 10^2$ & \phantom{-}0.0 & (-5.0,5.0) & Gauss(\phantom{-}0.0,1.0) \\
$\Delta z_l^5 \times 10^2$ & \phantom{-}0.0 & (-5.0,5.0) & Gauss(\phantom{-}0.0,1.0) \\
\end{tabular}
\end{ruledtabular}
\end{table}

\subsection{DES Y1 WL Analysis Code and Public Data}\label{subsec:Y1_code}

The versions of the code in the CosmoSIS \cite{Cosmosis_paper,Cosmosis_program} libraries and the parameters in CosmoSIS used to perform the WL analysis are the same as the ones used in the projection bias study for the 3x2pt analysis \cite{PRD_us}. The choice of library versions and parameter values was selected to be consistent with the results of DES Y1 WL analysis. For testing purposes, we used the data that was used in the DES Y1 weak lensing analysis.  For our study we used $200$ logarithmically spaced points ($N_{ell} = 200 $) to calculate the inner integrals in Equations \ref{eq:gal-clus} and \ref{eq:gal-shear}, and used Multinest \cite{Multinest_ref} with a tolerance value of $10^{-3}$, and $500$ live points to sample the posterior distributions. Chainconsumer \cite{Chainconsumer_ref} was used for processing the output of Multinest using a Gaussian Kernel Density Estimator algorithm and the width of the Gaussian Kernel was specified by setting the parameter KDE equal to $1.5$. The details about why these choices were made are discussed in Reference \cite{PRD_us}.
\begin{figure}[hbt]
%\begin{center}
\includegraphics[width=0.45\textwidth]{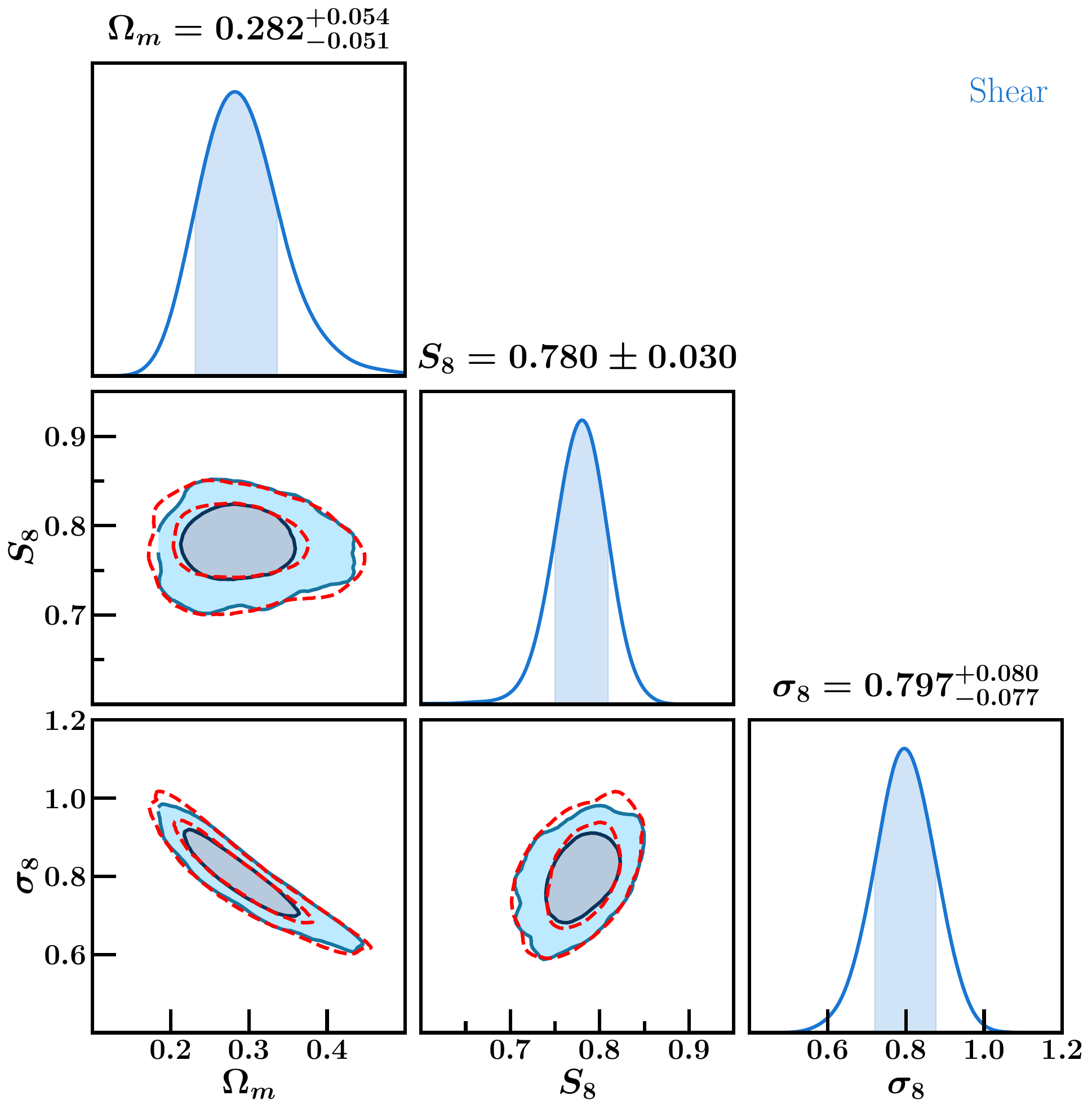}
\includegraphics[width=0.45\textwidth]{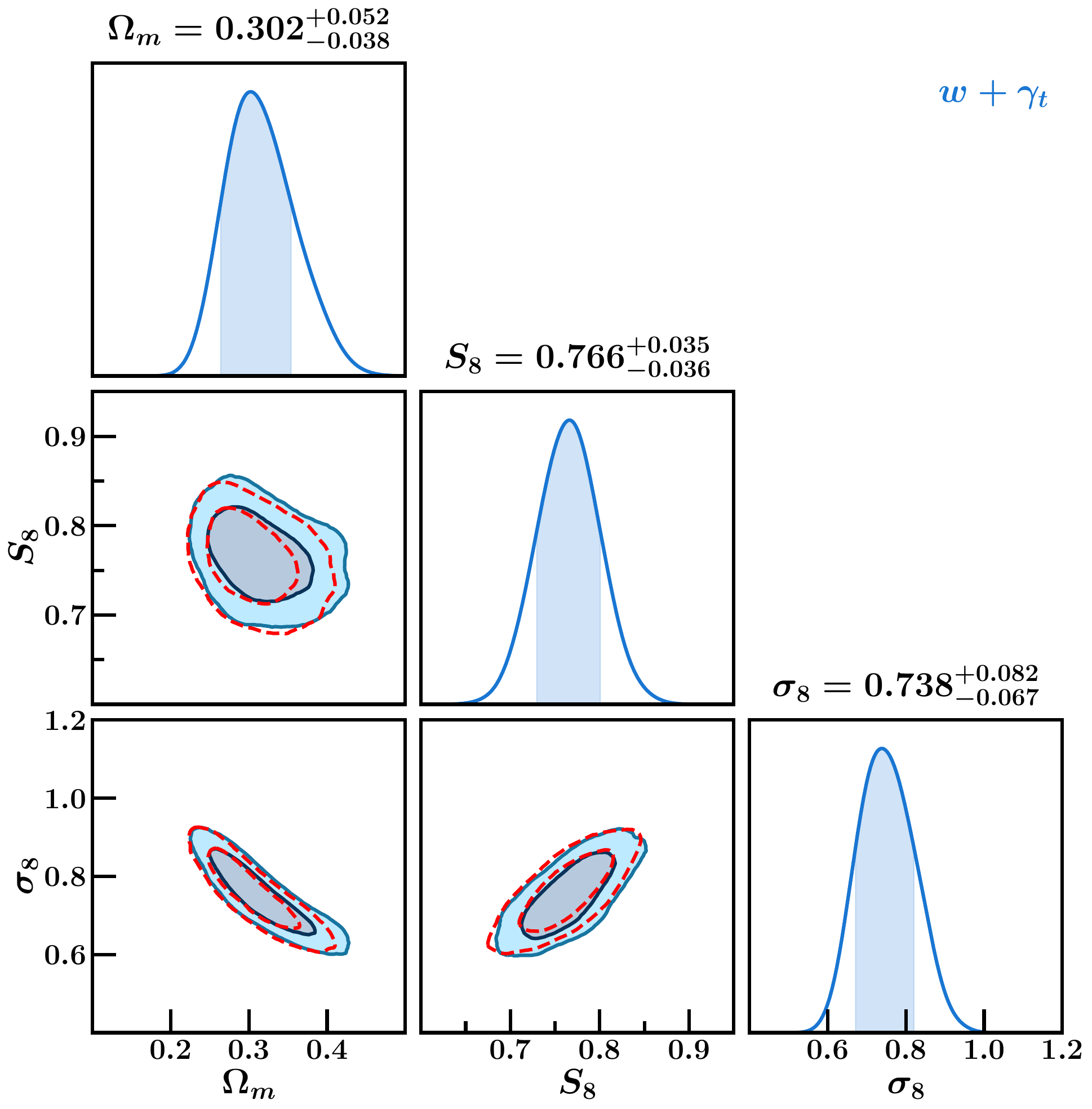}
\caption{Posterior distributions of $\Omega_m$, $\sigma_8$ and $S_8$ from the DES Y1 Shear analysis (top) and the DES Y1 $w + \gamma_{t}$ analysis (bottom). The solid (blue) lines in this figure show the results of our analysis of the DES Y1 WL data using the data and programs publicly available in CosmoSIS.  The dashed (red) line corresponds to the results published in DES Year 1 analysis (see Fig. 5 in reference \cite{DES_Y1_PRD}).}
\label{fig:Figure5_PRD_us}
%\end{center}
\end{figure}
\begin{table}[bht]
\caption{\label{tab:y1results} Comparison of the results of our analysis of the Y1 data (top two rows) with the results of the published DES Y1 WL analysis \cite{DES_Y1_PRD} (bottom two rows). }
\begin{ruledtabular}
\begin{tabular}{cccc}
$\Omega_{m}$& $S_{8}$& \multicolumn{2}{c}{\textrm{STUDY}}\\
\colrule\\
$0.282 ^{+0.054}_{-0.051}$ & $0.780 ^{+0.030}_{-0.030}$ & \multicolumn{2}{c}{Projection Bias 1x2pt}\\\\
$0.302 ^{+0.052}_{-0.038}$ & $0.766 ^{+0.035}_{-0.036}$ & \multicolumn{2}{c}{Projection Bias 2x2pt}\\
\\ \hline  \\
 $0.260^{+0.065}_{-0.037}$ & $0.782^{+0.027}_{-0.027}$ & \multicolumn{2}{c}{DES Y1 1x2pt} \\
 \\
 $0.288^{+0.045}_{-0.026}$ & $0.760^{+0.033}_{-0.030}$ & \multicolumn{2}{c}{DES Y1 2x2pt} \\
 \\
\end{tabular}
\end{ruledtabular}
\end{table}

Figure \ref{fig:Figure5_PRD_us} compares the results of our analysis of the publicly available DES Y1 WL data (in solid blue) with the DES Y1 WL results published in Figure 5 of reference \cite{DES_Y1_PRD} (in dashed red). These plots include one and two-dimensional posterior distributions for $\Omega_m$, $\sigma_8$ and $S_8$ corresponding to the DES 1x2pt Shear analysis (top plots), and the 2x2pt $w + \gamma_{t}$ analysis (bottom plots). We can see that the publicly available CosmoSIS code used in our studies reproduces the published results of the DES Y1 WL analysis very well. Table \ref{tab:y1results} shows the comparison of these results for the parameters $\Omega_m$ and $S_{8}$
($\sigma_{8}$ values were not published for the DES Y1 1x2pt and 2x2pt analysis).

As observed in the previous study \cite{PRD_us}, projection biases are a function of the position of the true values of the parameters with respect to their prior ranges.  So, to study the behavior of these biases, we have to vary the position of the true values of the parameters with respect to the prior intervals. While 16 parameters enter in the 1x2pt analysis, 26 parameters, 16 of which are used in 1x2pt analysis, enter in the 2x2pt analysis.  
For the 2x2pt (1x2pt) analysis, we used the same selected set (subset) of true or nominal values as the parameters that were used in the previous study of biases in the 3x2pt analysis. Tables \ref{tab:parameters} and \ref{tab:parameters2} list all the parameters used in the analysis. We studied the biases in the parameter space of the four cosmological parameters $h$, $\Omega_b$, $n_s$ and $\Omega_\nu h^2$ that have wide likelihood distributions and are correlated to $\Omega_m$, $\sigma_{8}$ and $S_{8}$. The rest of the parameters do not satisfy the conditions of 
wide likelihoods with respect to their prior intervals and a correlation to $\Omega_m$, $\sigma_{8}$ or $S_{8}$, and therefore do not contribute to the projection biases of these last three parameters.

A total of 84 different combinations (similar to the 3x2pt analysis study) of true values were used to study the biases produced by varying $h$, $\Omega_b$, $n_s$ and $\Omega_\nu h^2$. These true values are listed in Table \ref{tab:variables}. The three columns in the last four rows on the table show the first three combinations that were studied. Here, only the $\Omega_\nu h^2$ value is varied, while for $h$, $\Omega_b$, and $n_s$ we used the values obtained in our analysis of the DES Y1 WL data. The three $\Omega_\nu h^2$ values correspond to i) the minimum value allowed by neutrino oscillation experiments ($0.615\times10^{-3}$), ii) a value that is close to the middle of the prior range ($4.5\times10^{-3}$), and iii) a value close to the upper end of the prior range ($9.0\times10^{-3}$).  The first four rows in the same table show three values for $h$, $\Omega_b$, and $n_s$ selected as 25\%, 50\%, and 75\% of their respective prior ranges, plus the previously selected values of $\Omega_\nu h^2$.  All possible combinations $(3^{4})$ of the parameter values in the first four rows were also studied.
\begin{table}[bht]
\caption{\label{tab:variables}%
List of nominal values of the cosmological parameters labeled as ``varied" in Table \ref{tab:parameters}. These parameter values are used in the projection bias study.  All $81$ combinations of the values listed in the three columns of the first four rows, plus the three cases listed in the three columns of the last four rows were studied.}
\begin{ruledtabular}
\begin{tabular}{cccc}
\textrm{Parameter} & \multicolumn{3}{c}{\textrm{True value (Percent of prior range)}} \\
\hline
\multicolumn{4}{c}{\textrm{All $3^4$ combinations of the first four rows}} \\
\hline
 $h$ & 0.64 (25) & 0.73 (50) & 0.82 (75) \\
 $\Omega_{b}$ & 0.04 (25) & 0.05 (50) & 0.06 (75) \\
 $n_{s}$ & 0.92 (25) & 0.97 (50) & 1.02 (75) \\
 $\Omega_\nu h^2 \times 10^3$ & 0.615 (1.2) & 4.5 (42.1) & 9.0 (89.5) \\
\hline
\multicolumn{4}{c}{\textrm{plus the 3 cases in the following 3 columns}} \\
\hline
 $h$ & 0.692 (39) & 0.692 (39) & 0.692 (39) \\
 $\Omega_{b}$ & 0.0504 (51) & 0.0504 (51) & 0.0504 (51)  \\
 $n_{s}$ & 0.975 (52.5) & 0.975 (52.5) & 0.975 (52.5) \\
 $\Omega_\nu h^2 \times 10^3$ & 0.615 (1.2) & 4.5 (42.1) & 9.0 (89.5) \\
\end{tabular}
\end{ruledtabular}
\end{table}
%
%%%%%%%%%%%%%%%%%%%%%%%%%%%%%%%%%%%%%%%%%%%%%%%%%%%%%%%%%%%%%%%%%%%%%%
%%%%%%%%%%%%%%%%%%%% Our projection bias results  %%%%%%%%%%%%%%%%%%%%
%%%%%%%%%%%%%%%%%%%%%%%%%%%%%%%%%%%%%%%%%%%%%%%%%%%%%%%%%%%%%%%%%%%%%%

\section{Projection Biases}\label{sec:ProjectionBias}

\begin{figure}[hbt]
\includegraphics[width=0.45\textwidth]{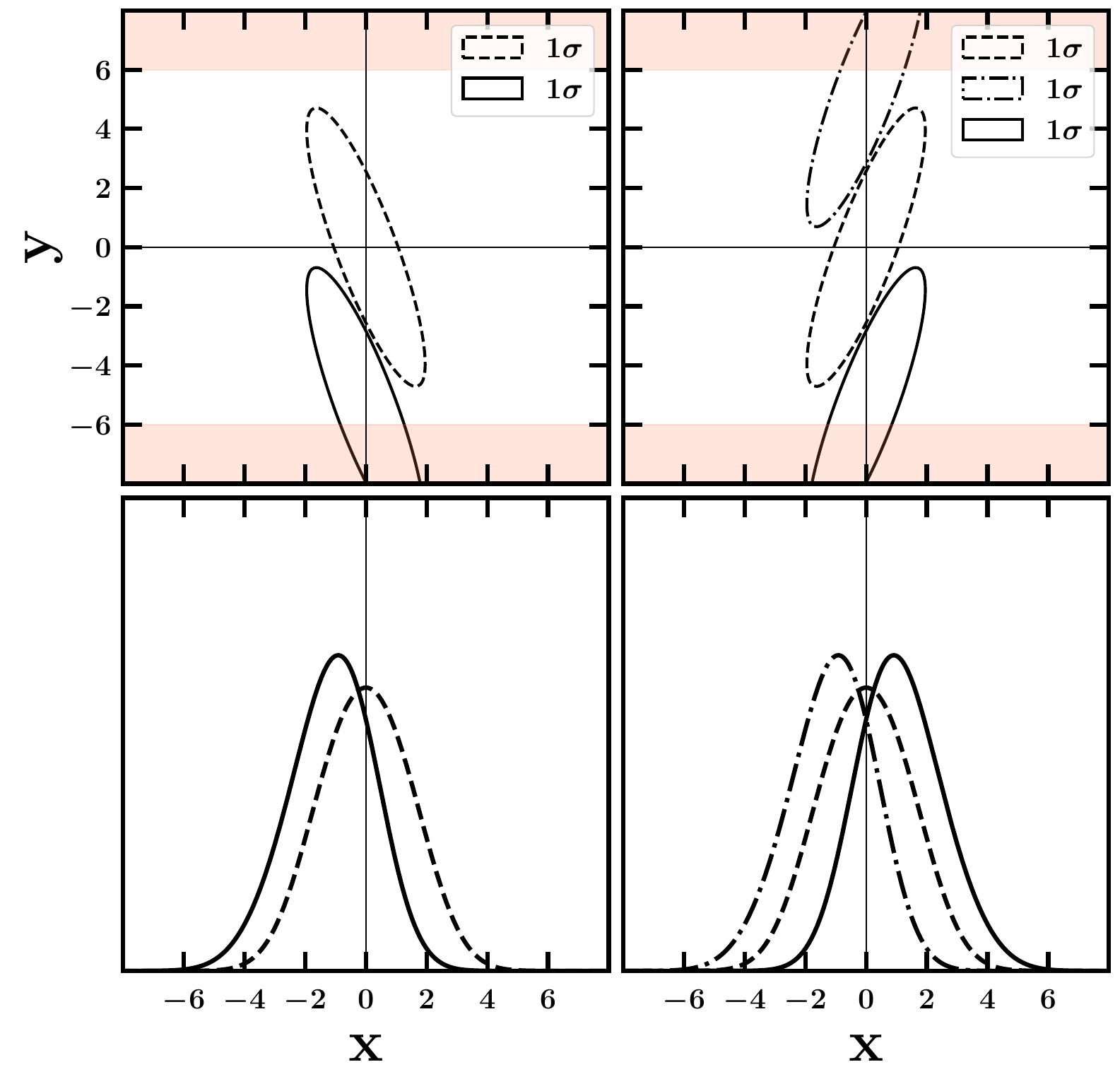}
\caption{Posterior distributions for a simple two dimensional Gaussian likelihood with parameters $X$ and $Y$. The top two plots show several two-dimensional ellipses corresponding to likelihood values of $\chi = 1$.  The center of the ellipses represent the true values of $X$ and $Y$. The unshaded area shows the prior range. The two bottom plots show the posteriors distribution in $X$ after marginalization over $Y$ - different line types correspond to the projections of the likelihood with the same line types in the plots on top. }
\label{fig:example}
\end{figure}

In this section, we will discuss the results of the projection bias study for the DES Y1 1x2pt and 2x2pt WL analysis, and compare them with the results from a similar study of the DES Y1 3x2pt WL analysis \cite{PRD_us}. To understand projection biases, consider the simple case of a two-dimensional Gaussian likelihood $\mathcal{L}(X,Y)$ (as in Eq. \ref{eq:2D-likelihood}) with parameters $X$ and $Y$, multiplied by a prior interval in $X$ that is very wide (like for $\Omega_m$, $\sigma_{8}$ and $S_{8}$ in the DES WL analysis), and a prior in $Y$ that is narrower than the width of the likelihood (like for $h$, $\Omega_b$, $n_s$, and $\Omega_\nu h^2$ in the DES WL analysis).  Several one sigma ellipses of the likelihood are shown in the top two plots in Figure \ref{fig:example}, where the un-shaded area in the plots show the range of the priors. The ellipses correspond to different ``true" (average) values of $Y$ and to opposite correlations between $X$ and $Y$.  The bottom two plots in the same figure show the posteriors in $X$ after marginalization (integration) over $Y$.  We see that projecting to the $X$-axis produces clear biases and that these ``projection" biases will depend on the correlation between the variables $X$ and $Y$, on the true value of $Y$, and on the width of the prior interval.  Note that the sign of the shift in the projected posterior peak depends on the correlations between $X$ and $Y$, and that these biases exist even in the presence of an unbiased likelihood (perfect theory).  For full details of this two-dimensional example, see section III A in reference \cite{PRD_us}.

\subsection{Bias Calculation Methodology}\label{subsec:quantifying}
The DES Y1 WL programs described in Section \ref{subsec:Y1_code} were used to calculate vectors with values of all three two-point correlations used in WL analysis (see Figures 2 and 3 in reference \cite{DES_Y1_PRD}).  These two-point correlation vectors are then used as input ``data" in the subsequent analysis, and are referred to as ``synthetic data vectors".  These synthetic data vectors are ``exact", in the sense that they were generated using the same theory and $\hat{n}^i_{g/\kappa}(z)$ distributions that are used in the analysis, therefore running the analysis code on these data vectors will give a $\chi^2$ value of zero.  This eliminates any systematic errors that could arise due to differences between the data and the theory that is used to analyze it.

To study the biases in the peaks of the posteriors, we generated synthetic data vectors using the 84 combinations of cosmological parameters given in Table \ref{tab:variables}, and described at the end of Section \ref{subsec:Y1_code}. These are the same synthetic data vectors used in reference \cite{PRD_us} (see Section III B).  The DES Y1 2x2pt and 1x2pt WL analysis codes were then used to analyze these ``data vectors".  The peaks of the posteriors (Peak) and the asymmetric errors $\sigma_+$ and $\sigma_-$, obtained from the 68.27\% credible intervals, were then used to calculate the pull defined as \cite{pull_reference,pull_LucDemortier}:
\begin{equation}
\left\{ \begin{array}{cl}
\mbox{pull} = (\textrm{Peak} - \textrm{True}) / \sigma_+ & \mbox{, if } \textrm{True} > \textrm{Peak} \\
\mbox{pull} = (\textrm{Peak} - \textrm{True}) / \sigma_- & \mbox{, if } \textrm{True} \le \textrm{Peak} \end{array}\right.
\label{eq:pull_definition}
\end{equation}
where True are the values of the cosmological parameters that were used to generate the synthetic data vectors.

To do a frequentist check of the credible intervals, and calculate coverage probabilities, we need to simulate performing the experiment many times while allowing for statistical fluctuations between experiments.  We did this by generating fluctuations using the analysis covariance matrix, adding the fluctuations to the synthetic data vectors, and then performing the 2x2pt or 1x2pt DES Y1 WL analysis.  This procedure eliminates any systematic errors that can arise from having a covariance matrix that doesn't accurately reflect the real fluctuations in the data.  We refer to this procedure of generating fluctuated data vectors, and then performing WL analysis on them, as doing ``ensemble tests".  Due to CPU time limitations, we only performed ensemble tests in five of the 84 cases for which we had synthetic data vectors.  In each of the five cases, 220 different experiments were performed.  Table \ref{tab:ensembles} lists the values of the cosmological parameters for which we performed ensemble tests.  These are the same cases used in the 3x2pt WL bias studies in reference \cite{PRD_us}.

\begin{table}[bht]
\caption{\label{tab:ensembles}%
List of the five cases for which the projections biases were studied using both ensemble tests and analyzing a single synthetic data vector. Each row has the set of nominal or true values of the cosmological parameters $h$, $\Omega_b$, $n_s$, and $\Omega_\nu h^2$. The numbers in parenthesis indicate the parameter value as a percent of the parameter's prior range.  The rest of the true values are given in Table \ref{tab:parameters}.}
\begin{ruledtabular}
\begin{tabular}{c|cccc}
 \# & $h$ & $\Omega_b$ & $n_s$ & $\Omega_\nu h^2 \times 10^3$ \\
\hline
 1 & 0.692 (39) & 0.0504 (51) & 0.975 (52.5) & 0.615 ($\phantom{1}1.2$) \\
 2 & 0.692 (39) & 0.0504 (51) & 0.975 (52.5) & $4.5\phantom{11}$ (42.1) \\
 3 & 0.692 (39) & 0.0504 (51) & 0.975 (52.5) & $9.0\phantom{00}$ (89.5) \\
 4 & $0.82\phantom{0}$ (75) & $0.04\phantom{00}$ (25) & $1.02\phantom{0}$ (75) & $4.5\phantom{11}$ (42.1) \\
 5 & $0.82\phantom{0}$ (75) & $0.04\phantom{00}$ (25) & $1.02\phantom{0}$ (75) & 0.615 ($\phantom{1}1.2$) \\
\end{tabular}
\end{ruledtabular}

\end{table}

From a histogram of the 220 pull values, or pull plot, one can extract the biases in the peak of the posterior, the biases in the credible intervals, and the coverage probabilities.  These studies are discussed in Section \ref{subsec:intervals}. The biases in the peaks of the posteriors and the widths of the pull plots were then used to estimate, as discussed in Appendix \ref{appendix:estimations}, the coverage probabilities for all 84 cases. 
In the following section, we discuss the biases in the peaks of the one-dimensional posteriors.

\subsection{Projection biases in the Posterior Peaks}\label{subsec:peaks}

Projection biases in general are best studied producing pull plots but, due to CPU limitations, it is usually not possible to explore a large parameter space this way.  In this section, we first show that for the biases in the peaks of the posteriors, pull plots and synthetic data vector analysis give very similar results. We then use the analysis of synthetic data vectors to explore a large parameter space.  

\begin{figure}[hbt]
\includegraphics[width=0.45\textwidth]{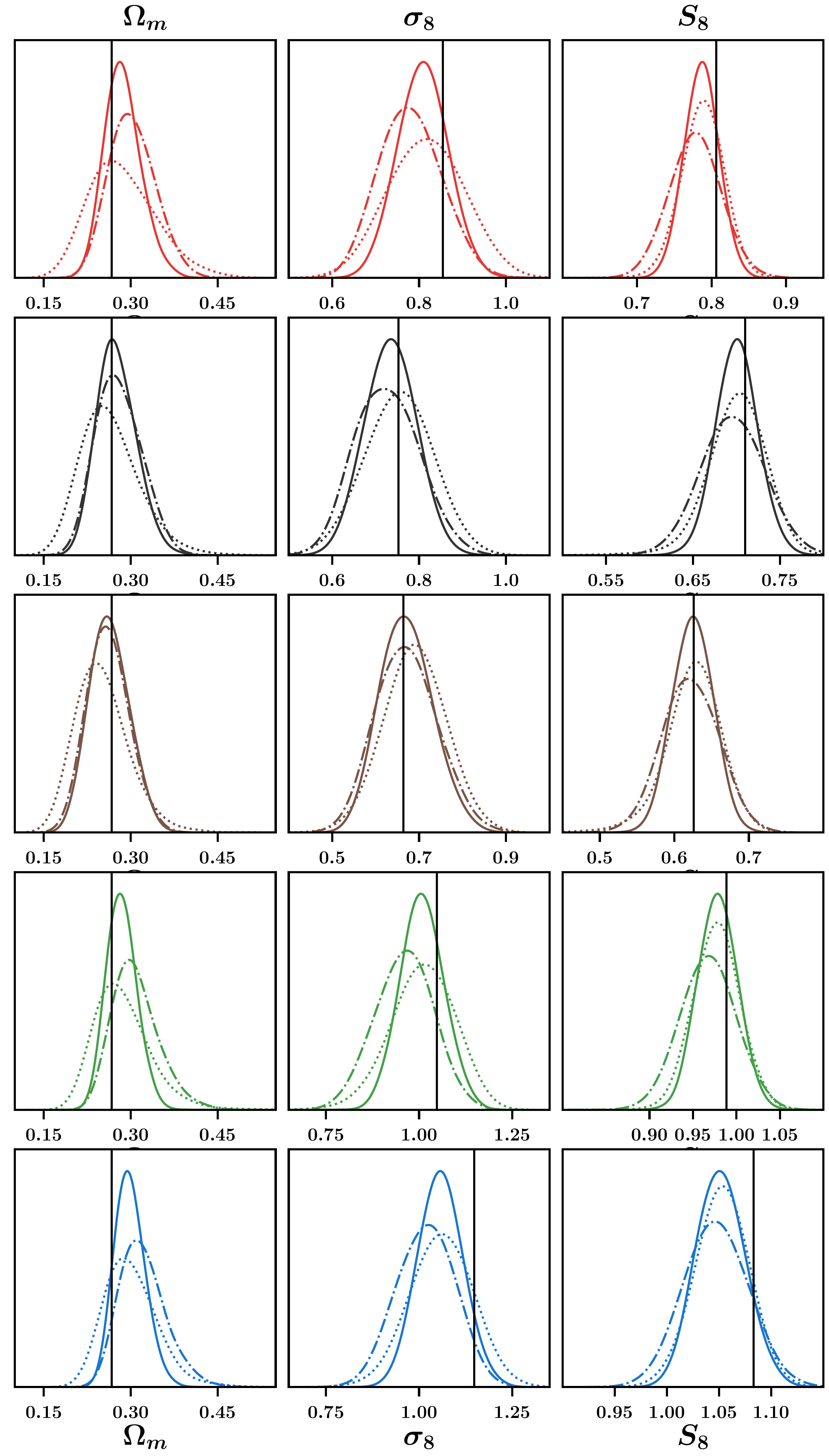}
\caption{Posterior distributions from the analysis of the synthetic data vectors created with the cosmological parameter values listed in Table \ref{tab:ensembles}. The posterior distributions given by the solid, dashed-dotted, and dotted lines correspond to the results from the 3x2pt, 2x2pt, and 1x2pt WL analysis respectively. The columns correspond to $\Omega_m$, $\sigma_8$ and $S_8$.  The rows labeled $1$ to $5$, are in the same vertical order as the one in Table \ref{tab:ensembles}.  The vertical solid lines correspond to the true, or input, value of the parameters.}
\label{fig:posteriors}
\end{figure}

Figure \ref{fig:posteriors} shows the posteriors obtained in the analysis of the five synthetic data vectors calculated using the cosmological parameters given in Table \ref{tab:ensembles}.  The posteriors corresponding to the 3x2pt, 2x2pt, and 1x2pt WL analyses are shown by the solid, dashed-dotted, and dotted lines respectively. The black solid lines show the true values of the parameters for each corresponding synthetic data vector.  Clear differences can be observed between the peaks of the posteriors and the true values of the parameters.

\begin{figure}[hbt]
\includegraphics[width=0.48\textwidth]{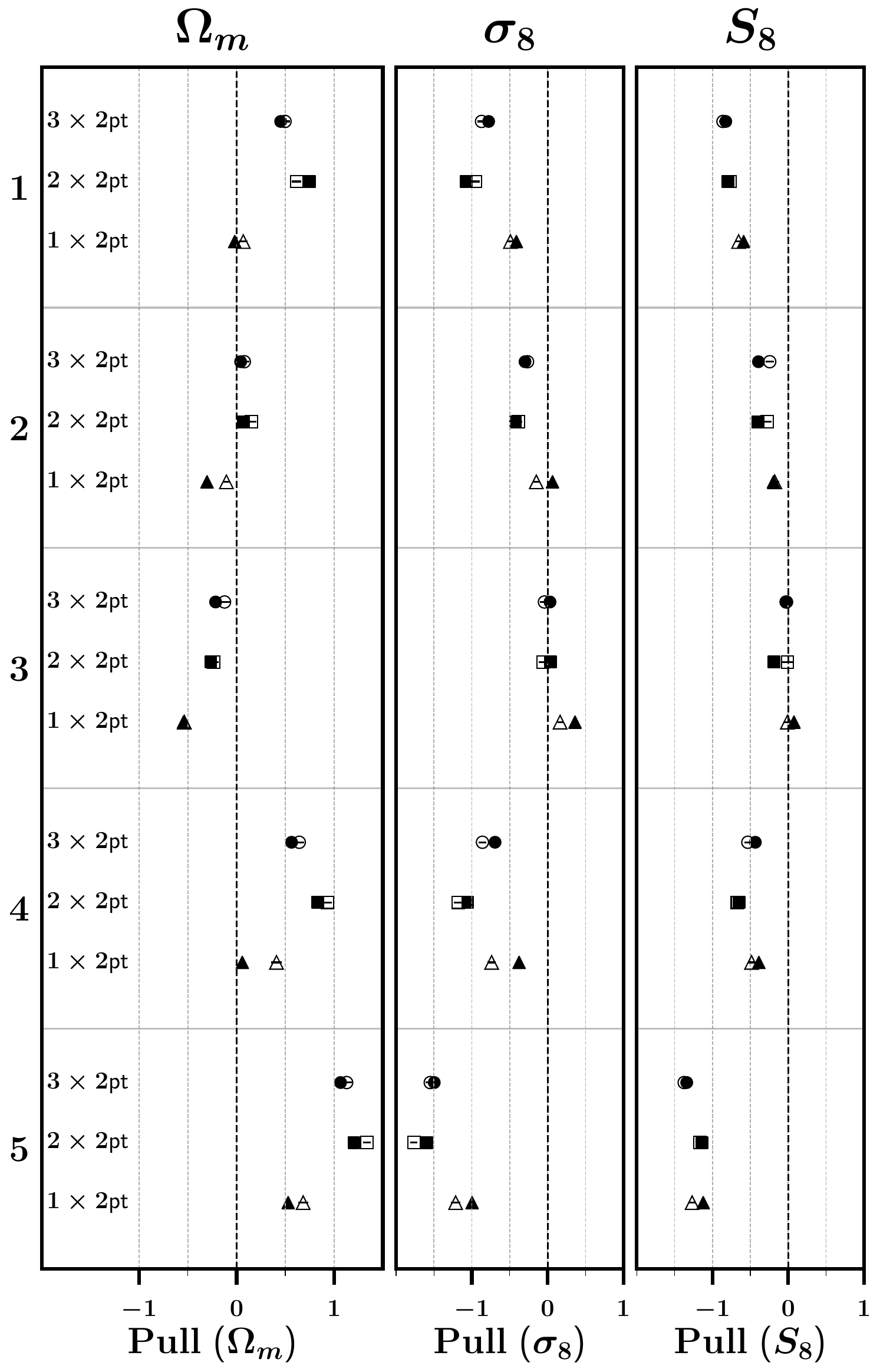}
\caption{The solid markers in this figure correspond to the pull, as in Equation \ref{eq:pull_definition}, constructed from the posteriors shown in Figure \ref{fig:posteriors}.  The open markers show $\bar{x}_p$, or the mean of the Gaussian fit to the pull histograms shown in Figure \ref{fig:pull_plot}.  In both cases, the circles, squares, and triangles represent the results from the 3x2pt, 2x2pt, and 1x2pt WL analysis respectively. The columns correspond to $\Omega_m$, $\sigma_8$, and $S_8$.  The vertical dashed lines correspond to zero bias, or the case when the peak of the marginalized posterior is equal to the true value.} 
\label{fig:posteriors-compare}
\end{figure}

Ensemble tests were also performed for the five cases shown in Figure \ref{fig:posteriors}, with the results of these studies discussed in Section \ref{subsec:intervals} and displayed in Figures \ref{fig:pull_plot} and \ref{fig:pull_inflate}.  The open markers in Figure \ref{fig:posteriors-compare} show $\bar{x}_p$, or the mean of the Gaussian fit to the pull histograms shown in Figure \ref{fig:pull_plot}, and the solid markers correspond to the pull (as in Equation \ref{eq:pull_definition}) constructed from the posteriors shown in Figure \ref{fig:posteriors}.  In both cases, the circles, squares, and triangles correspond to the 3x2pt, 2x2pt, and 1x2pt WL analysis respectively.  These results show that 1) biases of more than one $\sigma$ in the peaks of the posteriors exist for all three types of analyses, and 2) that the peaks of the posteriors in the synthetic data vector analysis agree very well with the peaks of the pull histograms.

\begin{figure*}[htb]
\includegraphics*[width=0.95\textwidth]{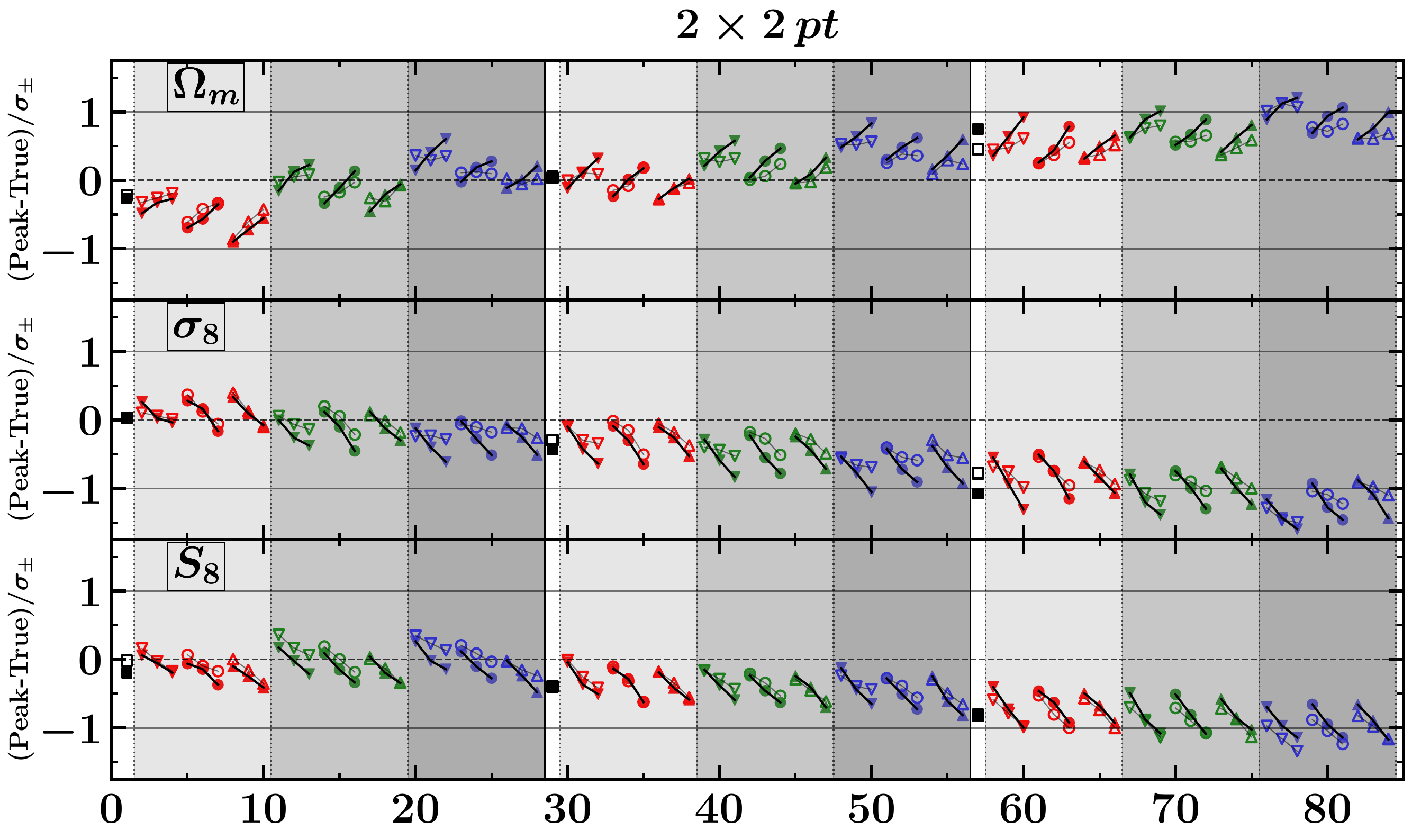}
\includegraphics*[width=0.95\textwidth]{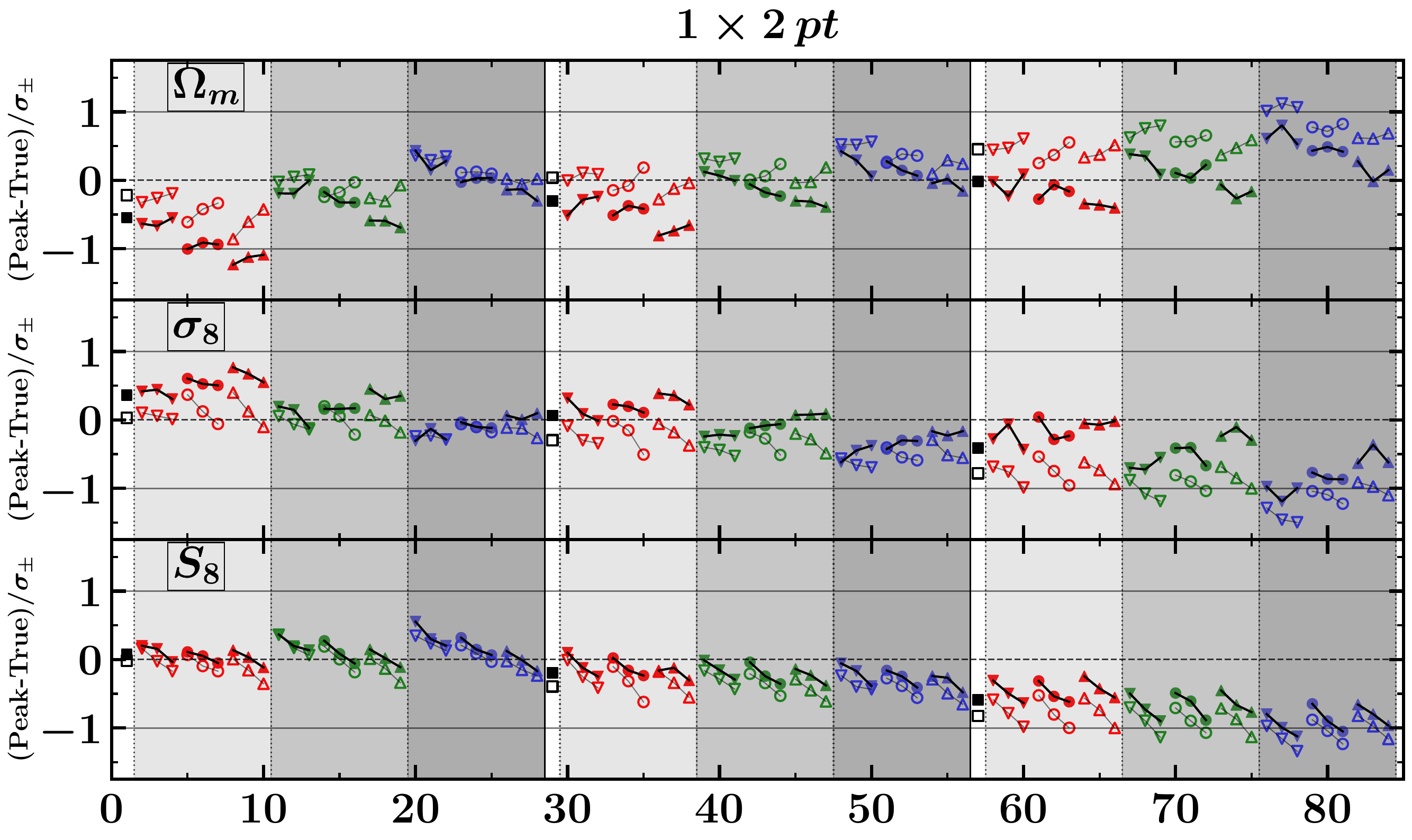}
%\begin{figure}[hbt]
%\includegraphics[width=0.48\textwidth]{Peak_bias.pdf}
\caption{Projection bias plots for the 2x2pt (top) and 1x2pt (bottom) WL analysis of the 84 combinations of parameter true values given in Table \ref{tab:variables}. The solid markers show the results of the 2x2pt and 1x2pt analysis. The open markers represent the results of the 3x2pt analysis of reference \cite{PRD_us}. The vertical scales show the pull obtained from the posterior distributions according to Equation \ref{eq:pull_definition}.  The horizontal scale shows the case number from 1 to 84.  Each plot is divided into three groups of 27 cases.  From left to right each group corresponds to values of $\Omega_\nu h^2$ equal to $89.5\%$, $42.1\%$, and $1.2\%$ of its prior range. Each group is further divided into light, medium, and dark grey shaded areas that correspond to different values of $h$. In each shaded region, the group of three points joined by a line represent cases with different values of $n_s$, and the downward triangles, circles, and upward triangles represent different values of $\Omega_b$. From left to right, the values of $h$, $n_s$, and $\Omega_b$ vary as the $25\%$, $50\%$, and $75\%$ of their prior ranges. Also from left to right, the three squares represent the cases listed in the last four rows of Table \ref{tab:variables}.}
\label{fig:peak_bias}
%\end{figure}
\end{figure*}
%%%%%start correcting here

Figure \ref{fig:peak_bias} shows the results of the projection biases in the peaks of the posteriors for $\Omega_m$, $\sigma_8$, and $S_8$, obtained as a result of the analysis of the 84 synthetic data vectors generated using the parameter values given in Table \ref{tab:variables}.  The solid markers show results for the 2x2pt analysis in the top plot and for the 1x2pt analysis in the bottom one.  For comparison, the results of the 3x2pt analysis, given in reference \cite{PRD_us}, are shown by the open markers.  The vertical axis in Figure \ref{fig:peak_bias} displays the pulls, as given in Equation \ref{eq:pull_definition}, and the horizontal axis shows the case numbers ordered as follows. Each plot is divided into three large groups of 27 cases.  From left to right, each group corresponds to values of $\Omega_\nu h^2$ equal to $89.5\%$, $42.1\%$, and $1.2\%$ of its prior range.  Each of these three groups is further divided into light, medium, and dark grey shaded areas which, from left to right, correspond to values of $h$ equal to $25\%$, $50\%$, and $75\%$ of its prior range.  In each of the shaded regions, the group of three points joined by a line represent, from left to right, the cases with values of $n_s$ equal to $25\%$, $50\%$, and $75\%$ of its prior range.  And finally, the downward triangles, circles, and upward triangles represent values of $\Omega_b$ equal to $25\%$, $50\%$, and $75\%$ of its prior range. From left to right, the three solid squares represent the cases listed in the last four rows of Table \ref{tab:variables}.

We can see, from the top plot in Figure \ref{fig:peak_bias}, that the projection biases for $\Omega_m$, $\sigma_{8}$, and $S_{8}$ in the 2x2pt WL analysis follow the biases in the 3x2pt analysis very closely.  We can also see that the projection biases in $\Omega_m$ increase with increasing values of $h$ and $n_s$, while the same biases decrease with increasing values of $\Omega_\nu h^2$ and $\Omega_b$.  This pattern is due to the correlations between $\Omega_m$ and the parameters with wide likelihoods, $h$, $n_s$, $\Omega_\nu h^2$, and $\Omega_b$.  This effect was briefly mentioned at the beginning of Section \ref{sec:ProjectionBias}, and it is explained in more detail in reference \cite{PRD_us}.  The biases in $\sigma_{8}$ and $S_{8}$ follow the opposite trend of the biases in $\Omega_{m}$, because of their opposite correlation with the parameters with wide likelihoods. The projection biases are minimal when the true values of the wide likelihood parameters are close to the middle of the prior intervals, and maximum when the true values are close to the edge of the intervals.

The overall trend of projection biases in the posterior peaks for the 1x2pt analysis is similar to that of the 3x2pt analysis, but the magnitude of the biases are different. This difference in magnitude may be due to the fact that, in the 1x2pt analysis, there are no galaxy bias parameters.  This reduces the number of free parameters from 26 in the 3x2pt and 2x2pt analyses to 16 in the 1x2pt analysis. In both the 1x2pt and 2x2pt analyses, the largest projection biases in $\Omega_{m}$, $\sigma_8$, and $S_8$ exceed the one $\sigma$ level.

\subsection{Biases in the 68.27\% Credible Intervals}\label{subsec:intervals}

\begin{figure*}[hbt]
\includegraphics[width=0.49\textwidth]{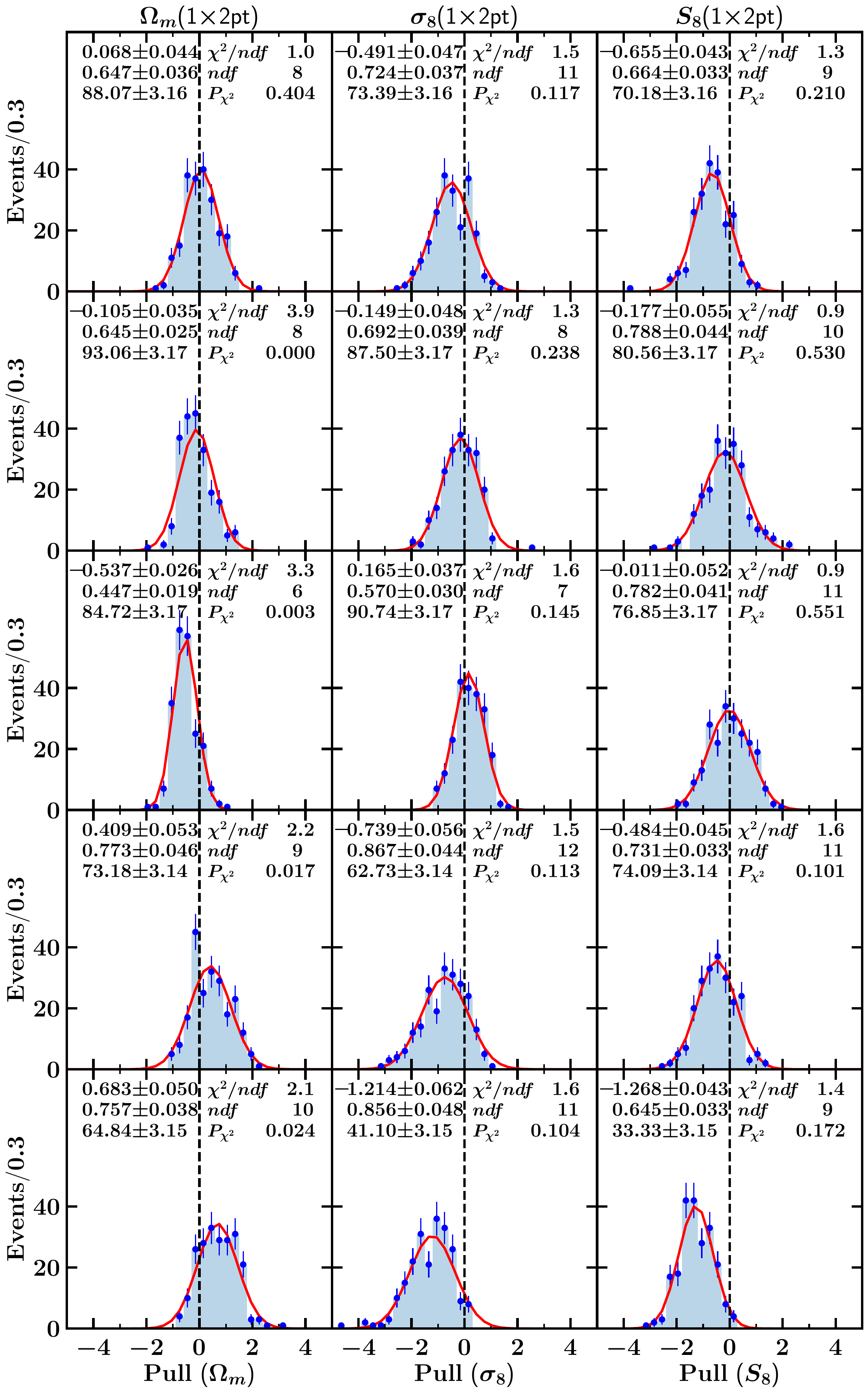}
\includegraphics[width=0.49\textwidth]{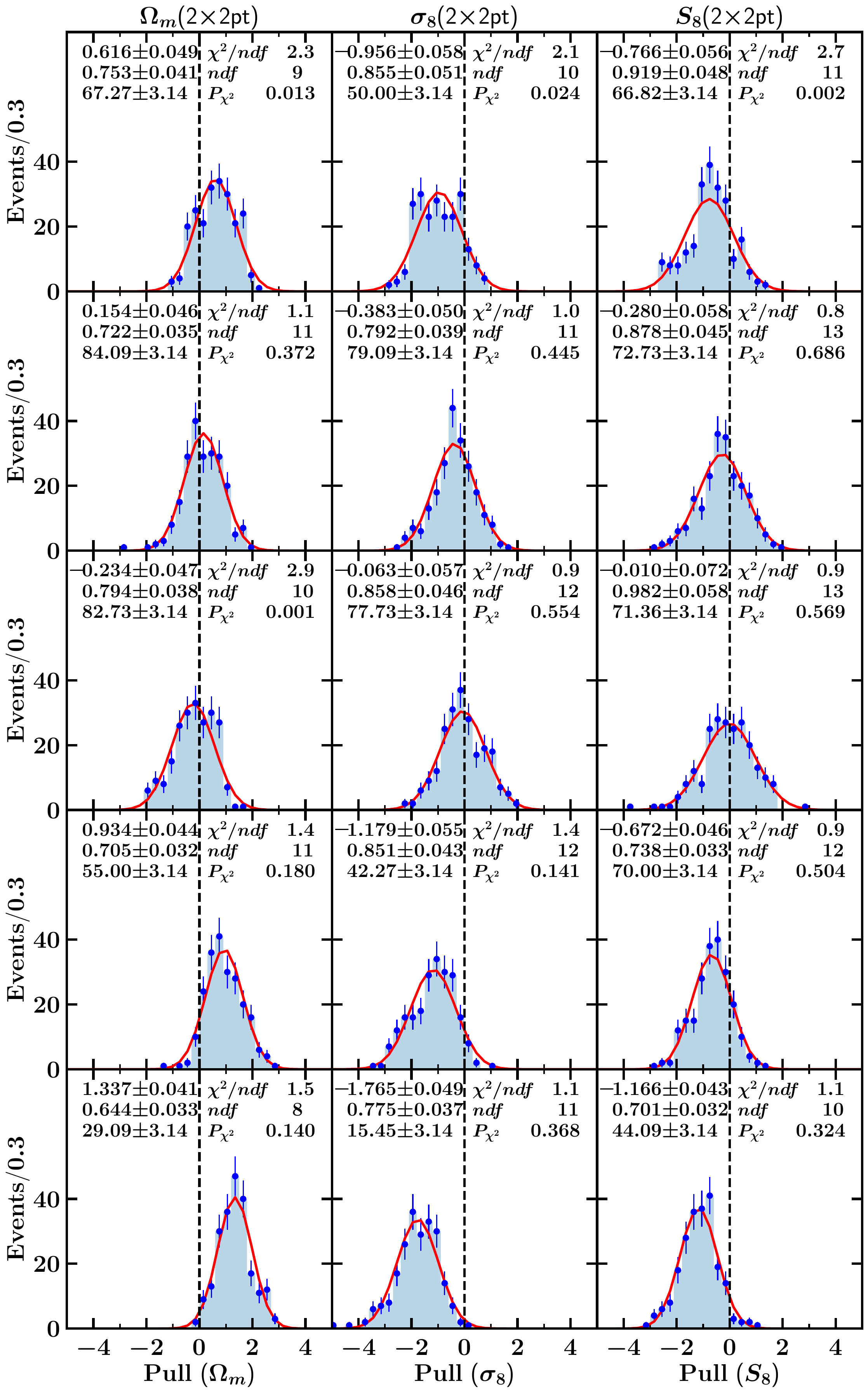}
\caption{Pull plots for the 1x2pt analysis on the left and for the 2x2pt analysis on the right for the five ensemble test cases listed in Table \ref{tab:ensembles}.  The horizontal axis in the histograms displays the pull values as calculated in Equation \ref{eq:pull_definition}.  The vertical order in the plots is the same as in Table \ref{tab:ensembles}.  The solid (blue) points with error bars show the number of entries in each bin with their binomial errors.  From top to bottom, the legends on the left side of the histograms show the mean $\bar{x}_p$ and sigma $\sigma_p$ of the Gaussian fits to the histogram (solid-red line), and the probability $P_{68}$ that the true value will lie inside the 68.27\% credible interval.  The right legends show $\chi^2/ndf$, the number of degrees of freedom $ndf$, and the $\chi^2$ probability of the fit.  }
\label{fig:pull_plot}
%\end{figure*}
%
%\begin{figure*}
\vspace{9mm}
\includegraphics[width=0.9\textwidth]{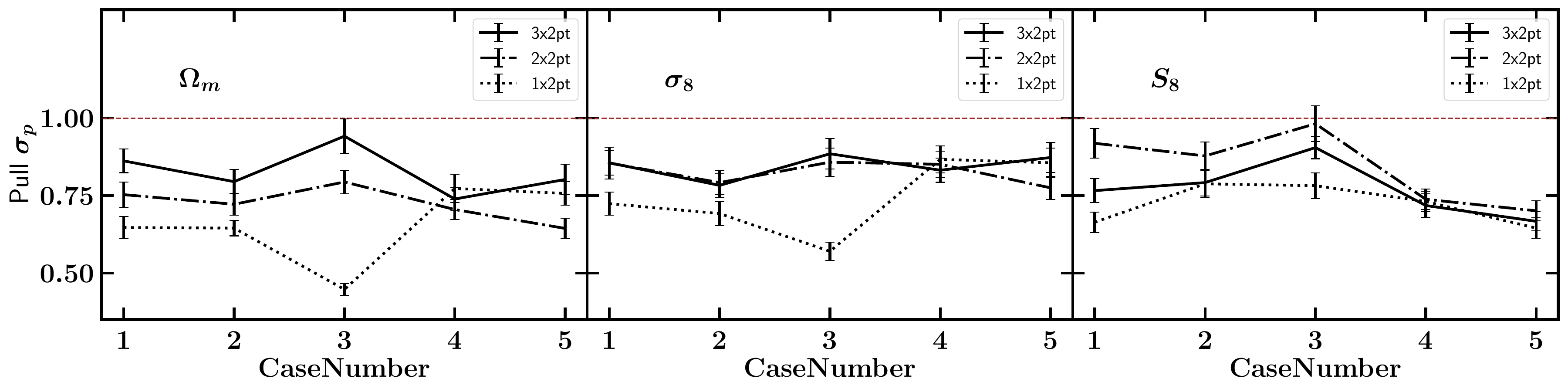}
\caption{Plots of $\sigma_p$, obtained from the Gaussian fit to the pull histograms, as a function of case number (vertical order in Figure \ref{fig:pull_plot}).  The 3x2pt, 2x2pt, and 1x2pt analysis are respectively connected by solid, dashed-dotted, and dotted lines.  In all cases $\sigma_p < 1$, indicating inflated credible intervals independent of the analysis and the values of $h$, $\Omega_b$, $n_s$, and $\Omega_\nu h^2$.  The 3x2pt analysis values come from Reference \cite{PRD_us}.}
\label{fig:pull_inflate}
\end{figure*}

In this section, we will cover the biases in the 68.27\% Credible Intervals. We will discuss the inflation of the DES Y1 WL credible intervals and also show the biases in the coverage probability.  To study these biases, we performed ensemble tests in five of the 84 cases shown in Figure \ref{fig:peak_bias}.  These five cases are listed in Table \ref{tab:ensembles}.  To reduce the errors in the bias calculations, each ensemble test consisted of 220 independent runs.  In each run random fluctuations, generated according to the covariance matrix, were added to the two-point correlation synthetic data vector.  Pulls were calculated for each of these runs according to Equation \ref{eq:pull_definition}.  Histograms of these pull values are shown in Figure \ref{fig:pull_plot}.
In an unbiased case, and using an infinite number of runs, these histograms, or pull plots, are expected to be a Gaussian distribution with mean value $\bar{x}_{p} = 0$ and standard deviation $\sigma_{p} = 1$ \cite{pull_LucDemortier}.  A value of $\bar{x}_{p} \ne 0$ indicates a bias in the peaks of the posteriors, and a value of $\sigma_{p} \ne 1$ indicates a bias in the credible intervals.  A value of $\sigma_{p} < 1$ indicates that the credible intervals are inflated, or larger than they need to be.  If $\sigma_{p} > 1$ then the credible intervals are too small.  The coverage probability $P_{68}$ is measured by counting the number of times the true value falls within the 68.27\% credible interval. For large statistics, shifting the posterior peak of every run by $-\bar{x}_{p}$ and multiplying the error by $\sigma_{p}$, will produce a pull plot with mean zero, standard deviation one, and $P_{68} = 68.27\%$.  In other words, these corrections transform a credible interval into a confidence interval.  The results of these corrections are discussed in \ref{subsec:case-for-conclusion}.

Figure \ref{fig:pull_plot} shows the pull plots for the 1x2pt WL analysis on the left and for the 2x2pt analysis on the right.  The vertical order in the pull plots corresponds to the same order shown in Table \ref{tab:ensembles} and in Figures \ref{fig:posteriors} and \ref{fig:posteriors-compare}.  The three columns in the right or left group of plots show the results for $\Omega_{m}$, $\sigma_8$ and $S_8$.  The solid (blue) dots with error bars in the histograms represent the number of entries, in bins of size 0.3, with their corresponding binomial error, and the solid (red) curves are the results of a Gaussian fit to the histogram.  The data used in the fits extend from -3$\sigma$ to +3$\sigma$ from the peak of the Gaussian.  The legends on the top left of each plot show, from top to bottom, the values of $\bar{x}_{p}$ and $\sigma_{p}$ from the Gaussian fits, and $P_{68}$ measured by counting as mentioned before. The legends on the top right show the value of $\chi^2/ndf$ and the number of degrees of freedom $ndf$, coming from the fit, and the $\chi^2$ probability $P_{\chi^2}$.  We checked the stability of the results with respect to the bin size.

The results from the pull plots show strong biases in the peaks of the posteriors, confirming what we observed in the study of synthetic data vectors.  The values of $\bar{x}_p$ were entered in Figure \ref{fig:posteriors-compare}, and show the similarity of the biases in the peaks of the posteriors obtained from the pull plots and the ones obtained from the synthetic data vector analysis.  Figure \ref{fig:pull_inflate} shows the values of $\sigma_p$ obtained from the fits to the pull plots.  We can clearly see that the errors obtained from the credible intervals are always inflated independent of the kind of WL analysis and of the true values of $h$, $\Omega_b$, $n_s$, and $\Omega_\nu h^2$.  Due to the strong biases in the posterior peaks, and the inflation of the credible intervals, the coverage probability, expected to be 68.27$\%$, actually ranges from values as low as $15.45\%$ to as high as $84.09\%$ in the 2x2pt analysis, and from values as low as $33.33\%$ to as high as $93.06\%$ in the 1x2pt analysis. 

Due to the large amount of CPU time required for doing ensemble tests, we only performed them for five of the 84 cases listed in Table \ref{tab:variables}.   For the other cases, we estimated the coverage probabilities using three reasonable assumptions. First, based on the good $\chi^2$ probabilities of the Gaussian fits in Figure \ref{fig:pull_plot}, we assumed that the pull histograms are Gaussian and calculated the coverage probabilities using Gaussian distributions. Second, for the mean $\bar{x}_p$ of the Gaussian we used the peaks of the posteriors calculated in the analysis of the synthetic data vectors.  The results plotted in Figure \ref{fig:posteriors-compare} show that this is a very good approximation. And finally, for the width of the Gaussian $\sigma_p$, we interpolated from the values obtained in the ensemble tests.  The details of these calculations, and the estimations of $P_{68}$, are given in Appendix \ref{appendix:estimations}.  The results show that there are strong biases in the coverage probability in essentially all 84 cases studied in this paper.

%%%%%%%%%%%%%%%%%%%%%%%%%%%%%%%%%%%%%%%%%%%%%%%%%%%%%%%%%%%%%%%%%%%%%%
%%%%%%%%%%%%%%%%%%%%%%%%%%%%% behavior  %%%%%%%%%%%%%%%%%%%%%%%%%%%%%%
%%%%%%%%%%%%%%%%%%%%%%%%%%%%%%%%%%%%%%%%%%%%%%%%%%%%%%%%%%%%%%%%%%%%%%
\section{Gains obtained BY correcting biases}\label{sec:behavior}

So far, the discussion has been centered on the study of projection biases in the posterior peaks and the $68.27\%$ credible intervals. In this section, we will discuss the advantages of correcting the biases and transforming the credible intervals into confidence intervals. We will show that by presenting the results in terms of confidence intervals, in addition to credible intervals, the errors could be reduced by an amount equivalent to increasing the amount of data by as much as a factor of three.

Using a simple example, in section \ref{subsec:ci-behavior-1} we will study how the credible intervals behave when the amount of data used in the analysis is increased by a factor of $n$. In section \ref{subsec:ci-behavior-2} we will show that, due to the use of priors that are narrower than the likelihood, the simple expectation that the credible intervals will scale as $1/\sqrt{n}$ (as the data increases by a factor of $n$) doesn't hold in the DES WL analysis. In section \ref{subsec:case-for-conclusion}, we will compare the error reduction obtained by correcting projection biases in credible intervals with the error reduction due to an increase in the amount of data. We will show that for the DES WL analysis, in many cases, the reduction obtained by correcting the biases is equivalent to an increase by a factor of three in the amount of data.
\subsection{Scaling of credible intervals, a simple example.}\label{subsec:ci-behavior-1}

\begin{figure}[htb]
\includegraphics*[width=0.48\textwidth]{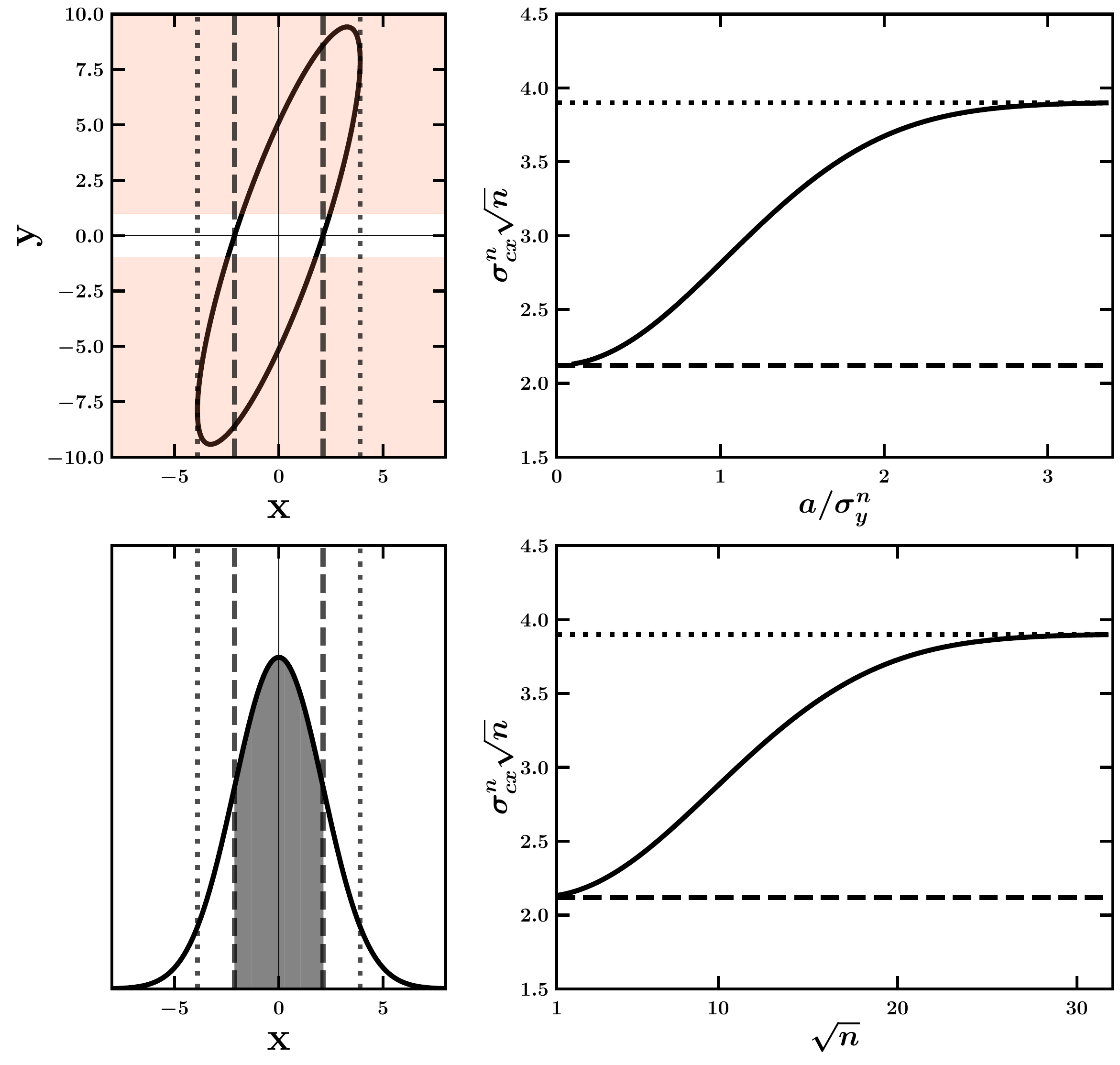}
\caption{The plots in this figure show the behavior of the 68.27\% credible interval with respect to an increase in statistics in a simple 2-dimensional example. The plots in the left column represent the starting, $n=1$, case in the study.  The ellipse in the top left plot shows the likelihood for $\chi^2 = 1$, the shaded area represents the region excluded by the prior. The bottom left plot shows the marginalized posterior $P(x)$, with the shaded area corresponding to the 68.27\% credible interval labeled as $2\, \sigma^{n=1}_{cx}$. The top right plot shows $\sigma_{cx}^{n} \sqrt{n}$ as a function of $a/\sigma_{y}^{n}$, or the ratio of prior to likelihood widths in the $y$ variable.  The bottom right plot is the same but as a function of $\sqrt{n}$. The dashed and dotted lines in all plots represent the lower and upper limits of $\sigma^n_{cx} \sqrt{n}$. See text for more details.}
\label{fig:2d-scale}
\end{figure}

To understand how credible intervals scale with the size of the data sample used in an analysis, we will start by studying this scaling in the simple example given in Ref. \cite{PRD_us}.  This simple two-dimensional example uses a normalized Gaussian likelihood, with correlated parameters $\mathbf{x}=(x,y)$, given by
\begin{equation}
\mathcal{L}(\mathbf{x}|\bar{\mathbf{x}}) = \frac{1}{2\pi\sigma_1 \sigma_2} \; e^{-\frac{1}{2} \chi^2} \;, \mbox{with} \; \chi^2 = \mathbf{u}^T C^{-1} \mathbf{u}.
\label{eq:2D-likelihood}
\end{equation}
and,
\begin{align}
    \mathbf{u} &=  (u,v) = [(x-\bar{x})/\sigma_x , (y-\bar{y})/\sigma_y], \\
    \sigma_x &= [(\cos\alpha / \sigma_1)^2 + (\sin\alpha/\sigma_2)^2]^{-1/2}, \\
    \sigma_y &= [(\sin\alpha/\sigma_1)^2 + (\cos\alpha / \sigma_2)^2]^{-1/2}.
    %C^{-1} &=  \left[ \begin{matrix} 1 & r \\  r & 1 \end{matrix} \right] \\
\end{align}
where $\sigma_1$, $\sigma_2$ and $\alpha$ are the semi-axes and rotation angle of the ellipse with $\chi^2 = 1$. $C$ is the covariance matrix with its inverse given as

\begin{align}
    C^{-1} &=  \left[ \begin{matrix} 1 & r \\  r & 1 \end{matrix} \right], \mbox{ with } \\
    r &=  \sigma_x \sigma_y \sin\alpha \cos\alpha \, (1/\sigma_1^2-1/\sigma_2^2).
\end{align}
The example uses an infinitely wide prior for $x$ and a finite prior for $y$ given as 
\begin{equation}
\pi(y) = \left\{ \begin{array}{cl}
1/2a & \mbox{if } -a \le y \le a \\ 0 & \mbox{otherwise.} \end{array}\right.
\label{eq:prior_y}
\end{equation}
The posterior, or normalized product of the likelihood and the prior is
\begin{equation}
P(\mathbf{x}) = \frac{\mathcal{L}(\mathbf{x}|\bar{\mathbf{x}}) \, \pi(y)}{\int \mathcal{L}(\mathbf{x}|\bar{\mathbf{x}}) \, \pi(y) \, d\mathbf{x}}
\label{eq:gen_posterior}
\end{equation}
The marginalization of the posterior over $y$ can be calculated exactly (see Equation $18$ in Ref. \cite{PRD_us}), and the result is
\begin{equation}
P(x) = \sqrt{ \frac{1-r^2}{2\pi \sigma_x^2}} \; e^{-\frac{1}{2} u^2 (1-r^2)} \left[\frac{\mbox{erf}(t_+) - \mbox{erf}(t_-)}{\mbox{erf}(z_+) - \mbox{erf}(z_-)} \right]
\label{eq:1D-projection}
\end{equation}
where $t_{\pm}$ and $z_{\pm}$ are 

\begin{align}
    t_{\pm} &= \frac{1}{\sqrt{2}}\left[ r u + \frac{\pm a - \bar{y}}{\sigma_{y}} \right], \\
    z_{\pm} &= \frac{\pm a - \bar{y}}{\sigma_{y}} \sqrt{\frac{1 - r^2}{2}}.
\end{align}
The credible interval in $x$ is calculated as the minimum interval containing 68.27\% of the area under the posterior distribution $P(x)$ given in Equation \ref{eq:1D-projection}.  In this section, we want to understand how this credible interval scales when the data sample is increased by a factor of $n$.  In order to do this, we divide the covariance matrix by $n$, $C^n = C/n$, which is equivalent to keeping $r$ constant and making the changes
\begin{equation}
    \sigma_{x}^{n} = \frac{\sigma_{x}}{\sqrt{n}}, \;\;\; \sigma_{y}^{n} = \frac{\sigma_{y}}{\sqrt{n}}
    \label{eq:scaling_n}
\end{equation}
For this example we select $a=1$, or $-1<y<1$, and we assume that the likelihood is centered at zero $(\bar{x},\bar{y}) = (0,0)$. For the semi-axes and rotation angle of the ellipse, we select $\sigma_{1} = 2$, $\sigma_{2} = 10$, and $\alpha = -20^\circ$, respectively.  The likelihood contour for $\chi^2 = 1$ is shown by the ellipse in the top left plot of Figure \ref{fig:2d-scale}. The marginalized posterior distribution $P(x)$ is shown in the bottom left plot in the same figure.  The shaded area in this plot represents the 68.27\% credible interval, which we denote as $2\, \sigma^n_{cx}$.  These credible intervals were calculated for values of $n$ ranging from $n=1$ to $1000$ with $n=1$ corresponding to the bottom left plot in Figure \ref{fig:2d-scale}.
The two plots on the right column of Figure \ref{fig:2d-scale} show how the credible interval in $x$ scales with the increase in statistics. The vertical axis in these two plots shows $\sigma_{cx}^{n} \sqrt{n}$.  The horizontal axis in the top plot displays the ratio of the width of the $y$-prior relative to the width of the likelihood along $y$, that is $a / \sigma^n_{y}$, and we will see that this gives a universal curve. In the bottom plot, the horizontal axis is $\sqrt{n}$.  As expected from Equation \ref{eq:scaling_n}, in the absence of priors, these two plots should display a straight line which they clearly don't.  The lower limit (dashed line) represents the case in which the prior is so narrow that $y$ is essentially set to zero, giving a value of $\sigma_x = 2.12$.  The upper limit (dotted line) represents the case in which the prior is infinitely wider than the likelihood, giving a value of $\sigma_x / \sqrt{1-r^2} = 3.90$.  These two limits can be clearly seen by the dashed and dotted lines shown in the left column plots in Figure \ref{fig:2d-scale}.  In these two limits, the credible intervals scale with the expected $1/\sqrt{n}$ behavior but, between the limits, the gain obtained with an increase in statistics slows down. This is because, as the statistics increase, the prior becomes less and less effective.
\begin{figure}[htb]
\includegraphics*[width=0.48\textwidth]{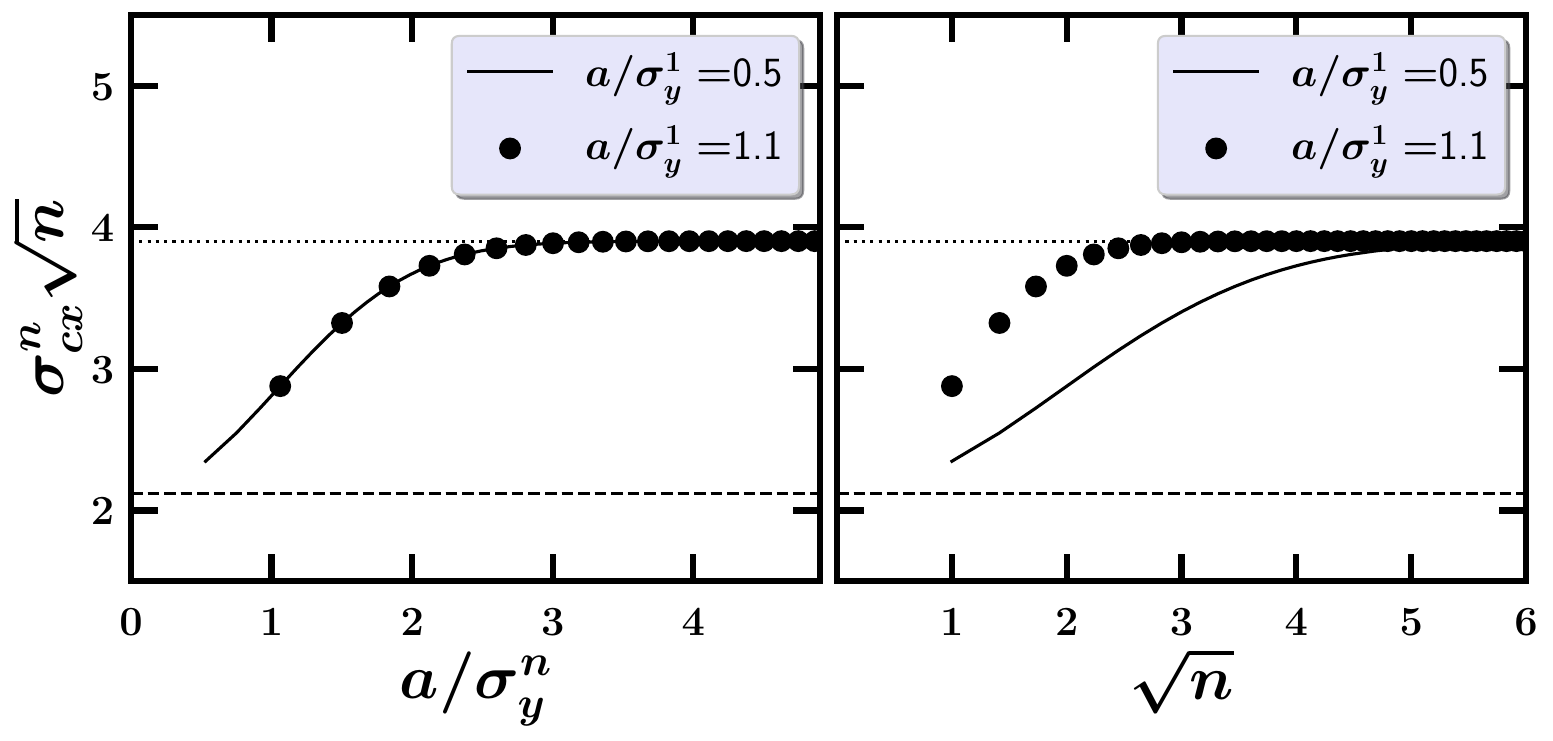}
\caption{This figure shows the results of a study similar to the one shown in Figure \ref{fig:2d-scale}, but with the ratio $a / \sigma_{y}^{n=1}$ taking the values $0.5$ for the solid line, and $1.1$ for the solid circles.  The curve on the left plot has a universal shape.}
\label{fig:2d-scale-2}
\end{figure}

Figure \ref{fig:2d-scale-2} shows the behavior of the credible interval for two initial ratios ($n=1$) of prior to likelihood widths.  The values are shown in the plot's insets. Even though we only show plots for two initial values of $a / \sigma_{y}^1$, the curve in the left plot is universal in the sense that it is independent of $a$ and $\sigma_{y}^1$.  On the right plot, the same values are plotted as a function of $\sqrt{n}$ which now shows that the two cases are different.  In practice, it is difficult to obtain the value of the likelihood for infinitely wide priors. For example, the likelihood cannot be calculated for negative neutrino masses.  Therefore, in the following section, when we study the scaling of credible intervals in the DES WL analysis, we will plot the results as a function of $\sqrt{n}$.  This will not give us a universal curve but will give us a curve with the same shape. 
%
%
%
%%%%%%%%%%%%%%%%%%%%%%%%%%%%%%%%%%%%%%%%%%%%%%%%%%%%%%%%%%%%%%%%%%%%%%
\subsection{Scaling of credible intervals in the DES Y1 WL analysis}\label{subsec:ci-behavior-2}
%%%%%%%%%%%%%%%%%%%%%%%%%%%%%%%%%%%%%%%%%%%%%%%%%%%%%%%%%%%%%%%%%%%%%%
%
In this section, we study how the credible intervals for $\Omega_m$, $\sigma_8$, and $S_8$, three key parameters in the DES Y1 WL studies, scale with a simulated increase of the amount of data (statistics) used in the analysis. An increase in statistics by a factor of 3, 6, and 9 times the data available in the Y1 analysis was simulated by respectively dividing the Y1 covariance matrix by 3, 6, and 9, and then redoing the analysis.  This simulated increase in statistics was applied to the analysis of all 84 synthetic data vectors used in the projection bias studies in previous sections.  The quantity $\sigma_a = (\sigma_+ + \sigma_-)/2$ was calculated from the posterior distributions obtained in the scaled analysis of each of the synthetic data vectors, and then, for each value of $n$, $\sigma^n$ was calculated averaging the values of $\sigma_a$ for all 84 synthetic data vectors.
\begin{figure}[htb]
\includegraphics*[width=0.48\textwidth]{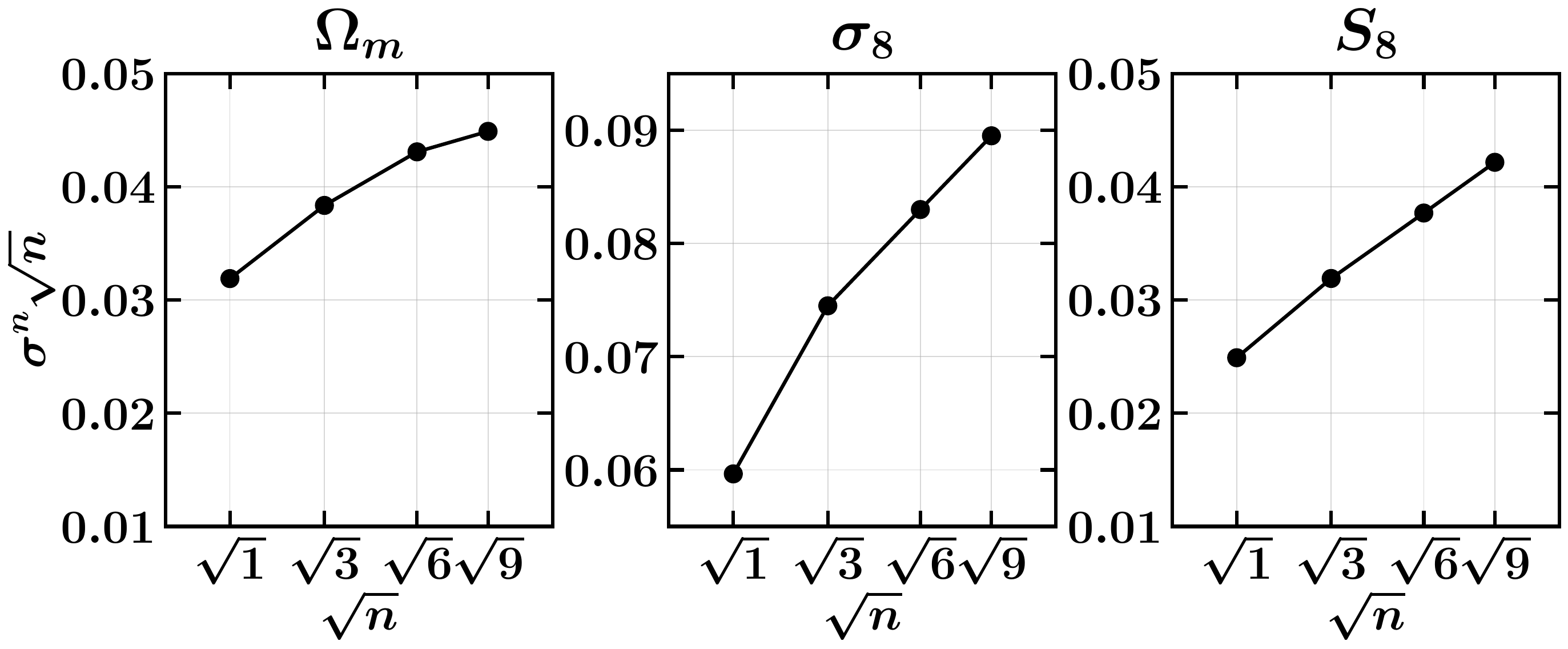}
\caption{These plots show the value of $\sigma^n \sqrt{n}$ as a function of $\sqrt{n}$ for $\Omega_m$, $\sigma_8$ and $S_8$.  The horizontal axis values of $\sqrt{1}, \sqrt{3}, \sqrt{6}, \sqrt{9}$ represent an increase in the amount of data available in the DES Y1 analysis of 1, 3, 6, and 9 times respectively.  The value of $\sigma^n$ in each data point is an average of the credible interval widths obtained in the 3x2pt analysis of the 84 synthetic data vectors used in previous sections.  See text for more details.}
\label{fig:y1-scale}
\end{figure}
The results of the study for the 3x2pt WL analysis, using the Y1 programs, are shown in Figure \ref{fig:y1-scale}. We see that the credible intervals have the same behavior observed in the simple example studied in the previous section. These results indicate that, after a factor of nine increase in the amount of Y1 data, the $\Omega_{m}$ credible interval approaches the upper plateau where errors scale as $1/\sqrt{n}$ while the errors for $\sigma_8$ and $S_8$ are still far from that plateau.
%
%%%%%%%%%%%%%%%%%%%%%%%%%%%%%%%%%%%%%%%%%%%%%%%%%%%%%%%%%%%%%%%%%%%%%%
\subsection{Bias corrections vs increase in statistics.}\label{subsec:case-for-conclusion}
%%%%%%%%%%%%%%%%%%%%%%%%%%%%%%%%%%%%%%%%%%%%%%%%%%%%%%%%%%%%%%%%%%%%%%
%
%
\begin{figure*}[htb!]
\includegraphics*[width=\textwidth]{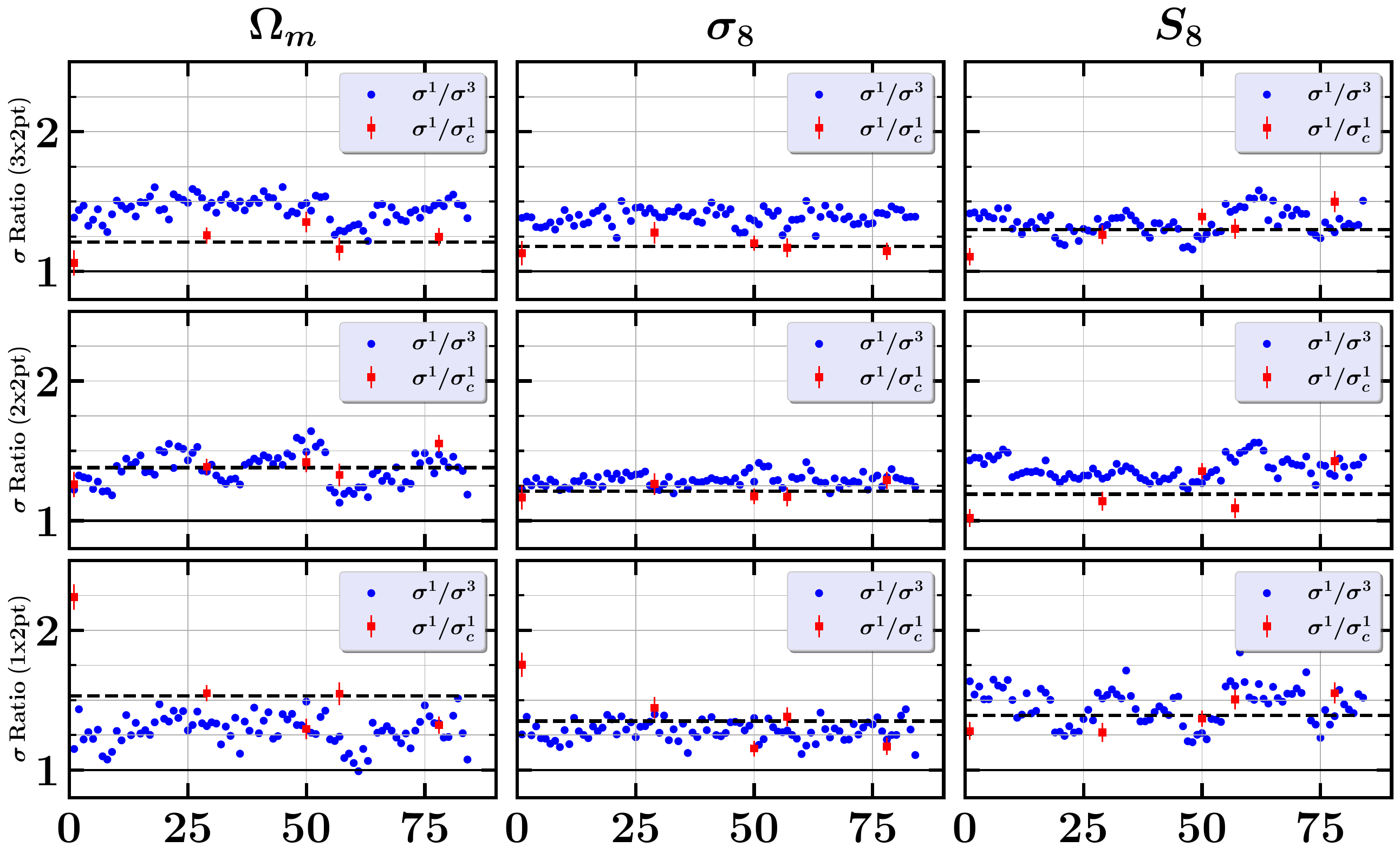}
\caption{This figure compares the gain obtained transforming the credible intervals of $\Omega_m$, $\sigma_8$ and $S_8$ into confidence intervals ($\sigma^{1}/\sigma^{1}_{c}$, red squares) with the gain obtained simulating an analysis with three times the Y1 data ($\sigma^{1}/\sigma^{3}$, blue circles). The vertical axis represents either $\sigma^1/\sigma^{1}_c$ or $\sigma^1/\sigma^3$, and the horizontal axis represents the case number as displayed in Figure \ref{fig:peak_bias}. The red squares with error bars represent the corrected credible intervals for the five cases for which we performed ensemble tests. The blue circles represent the gain in the credible intervals from the analysis of the 84 synthetic data vectors used in previous sections after the covariance matrix was divided by a factor of three. The dashed line represents the average of the red squares. The rows from top to bottom correspond to the results from the 3x2pt, 2x2pt, and 1x2pt analysis respectively.}
\label{fig:y1-scale_all}
\end{figure*}
As discussed in \ref{subsec:intervals}, we use $\bar{x}_p$ and $\sigma_{p}$ to transform the credible intervals into confidence intervals. The values of the width of the pull plots $\sigma_p$, displayed in Figures \ref{fig:pull_plot} and \ref{fig:pull_inflate}, clearly show that the credible intervals in the DES Y1 WL analysis are inflated by about 20\%, or $\sigma_p \approx 0.8$.  The use of $\sigma_p$ to transform the WL analysis credible intervals into approximately 20\% smaller confidence intervals can naively be compared to the gain obtained by increasing the data sample, leading one to conclude that it is equivalent to an increase in the amount of data of $1/0.8^2 = 1.56$.  As we saw in Sections \ref{subsec:ci-behavior-1} and \ref{subsec:ci-behavior-2}, due to the effect of priors, the scaling doesn't follow the simple $1/\sqrt{n}$ rule.  In this section we will do a detailed comparison of the two effects: the gain $1/\sigma_p$ produced by transforming the Credible Intervals into Confidence Intervals, and the effect of increasing the data sample by a factor of three, which is the gain in statistics between the Y1 and Y3 DES WL analysis.

If $\sigma^1$ is the error obtained from the Y1 WL analysis credible intervals and $\sigma^{1}_c$ is the error of the corresponding confidence interval, then $\sigma^1/\sigma^{1}_c = 1/\sigma_p$. The red squares with error bars in Figure \ref{fig:y1-scale_all} show the values of $\sigma^1/\sigma^{1}_c$ for the five cases for which we performed ensemble tests.  These cases are listed in Table \ref{tab:ensembles} and the results of the analysis are discussed in Section \ref{subsec:intervals}.  The case numbers given in the horizontal axis follow the same nomenclature as the one used in Figure \ref{fig:peak_bias}.  The dashed line in Figure \ref{fig:y1-scale_all} is the average of the red square values which, together with their maximum and minimum values, are also given in Table \ref{tab:Gains}.

\begin{table}[bht]
\caption{\label{tab:Gains}%
Gains obtained after correcting the $\Omega_{m}$, $\sigma_{8}$, and $S_{8}$ credible intervals for the 3x2pt, 2x2pt and 1x2pt WL analysis (red squares in Figure \ref{fig:y1-scale_all}). Max, Min, and Avg correspond to the maximum, minimum, and average gains for the five ensemble test cases listed in Table \ref{tab:ensembles}.}
\begin{ruledtabular}
\begin{tabular}{ccccc}
Analysis & Gains & $\Omega_{m}$ & $\sigma_{8}$ & $S_{8}$\\
\hline
 \multirow{3}{*}{3x2pt} & Max & 1.35 & 1.28 & 1.50 \\
                        & Min & 1.06 & 1.13 & 1.11 \\
                        & Avg & 1.21 & 1.18 & 1.30 \\
\hline
\multirow{3}{*}{2x2pt} & Max & 1.55 & 1.29 & 1.43 \\
                        & Min & 1.26 & 1.17 & 1.02 \\
                        & Avg & 1.38 & 1.21 & 1.19 \\
\hline
\multirow{3}{*}{1x2pt} & Max & 2.24 & 1.75 & 1.55 \\
                        & Min & 1.29 & 1.15 & 1.27 \\
                        & Avg & 1.53 & 1.35 & 1.39 \\
\end{tabular}
\end{ruledtabular}
\end{table}

The blue dots in Figure \ref{fig:y1-scale_all} correspond to the ratio $\sigma^1/\sigma^3$ where $\sigma^3$ is the error obtained from the credible intervals after the covariance matrix was divided by a factor of three.  The blue dots then show the gain obtained by simulating an increase in the data size by a factor of three for all the cases listed in Table \ref{tab:variables}.  Again, the horizontal axis uses the same nomenclature that was used in Figure \ref{fig:peak_bias}.  It is clear that on average, in six of the nine cases shown in Figure \ref{fig:y1-scale_all}, the corrections that transform the credible intervals into confidence intervals produce gains similar to or greater than the error reduction obtained in going from the DES Y1 to Y3 WL analysis, or to increasing the amount of data by a factor of three.

%%%%%%%%%%%%%%%%%%%%%%%%%%%%%%%%%%%%%%%%%%%%%%%%%%%%%%%%%%%%%%%%%%%%%%
%%%%%%%%%%%%%%%%%%%%%%% Increased statistics  %%%%%%%%%%%%%%%%%%%%%%%%
%%%%%%%%%%%%%%%%%%%%%%%%%%%%%%%%%%%%%%%%%%%%%%%%%%%%%%%%%%%%%%%%%%%%%%
%
\section{Extrapolating to Larger Data Sets}\label{sec:LargerStatistics}
%\begin{figure*}[hbt]
%\includegraphics*[width=\textwidth]{Plots/plot-all-2x2pt-cov3.pdf}
%\includegraphics*[width=\textwidth]{Plots/plot-all-1x2pt-cov3.pdf}
%\caption{Estimation of the Projection biases in the peaks of the posteriors for the parameters $\Omega_{m}$, $\sigma_{8}$ and $S_{8}$ measured in 1x2pt analysis (on top) and 2x2pt analysis (on bottom) for a simulated increase of statistical power by reducing the covariance matrix of the fit to two-point correlations functions by a factor of 3. The plots follow the same style of plots shown in Figure \ref{fig:peak_bias}.  The factor of three was selected to simulate the recently released DES Y3 Weak Lensing analysis.}
%\label{fig:peak_bias_3}
%\end{figure*}
%
%\begin{figure*}[hbt]
%\includegraphics*[width=\textwidth]{Plots/plot-all-2x2pt-cov9.pdf}
%\includegraphics*[width=\textwidth]{Plots/plot-all-1x2pt-cov9.pdf}
%\caption{Estimation of the Projection biases in the peaks of the posteriors for the parameters $\Omega_{m}$, $\sigma_{8}$ and $S_{8}$ measured in 1x2pt analysis (on top) and 2x2pt analysis (on bottom) for a simulated increase of statistical power by reducing the covariance matrix of the fit to two-point correlations functions by a factor of nine. The plots follow the same style of plots shown in Figure \ref{fig:peak_bias}.  The factor of nine was selected to simulate an optimistic increase in statistical power due to the full analysis of six years of DES data.}
%\label{fig:peak_bias_9}
%\end{figure*}
\begin{figure*}[hbt]
\includegraphics*[width=0.48\textwidth]{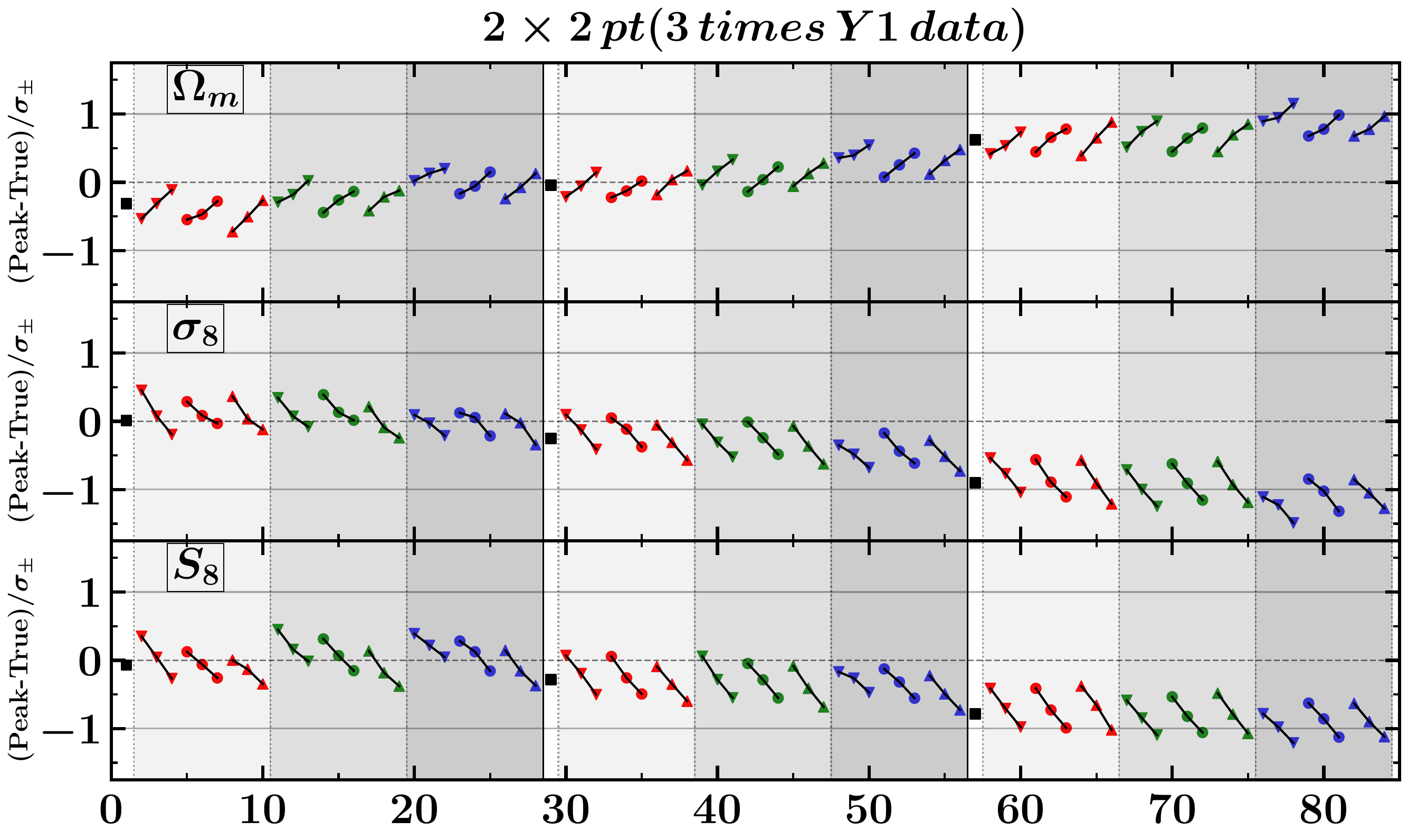}
\includegraphics*[width=0.48\textwidth]{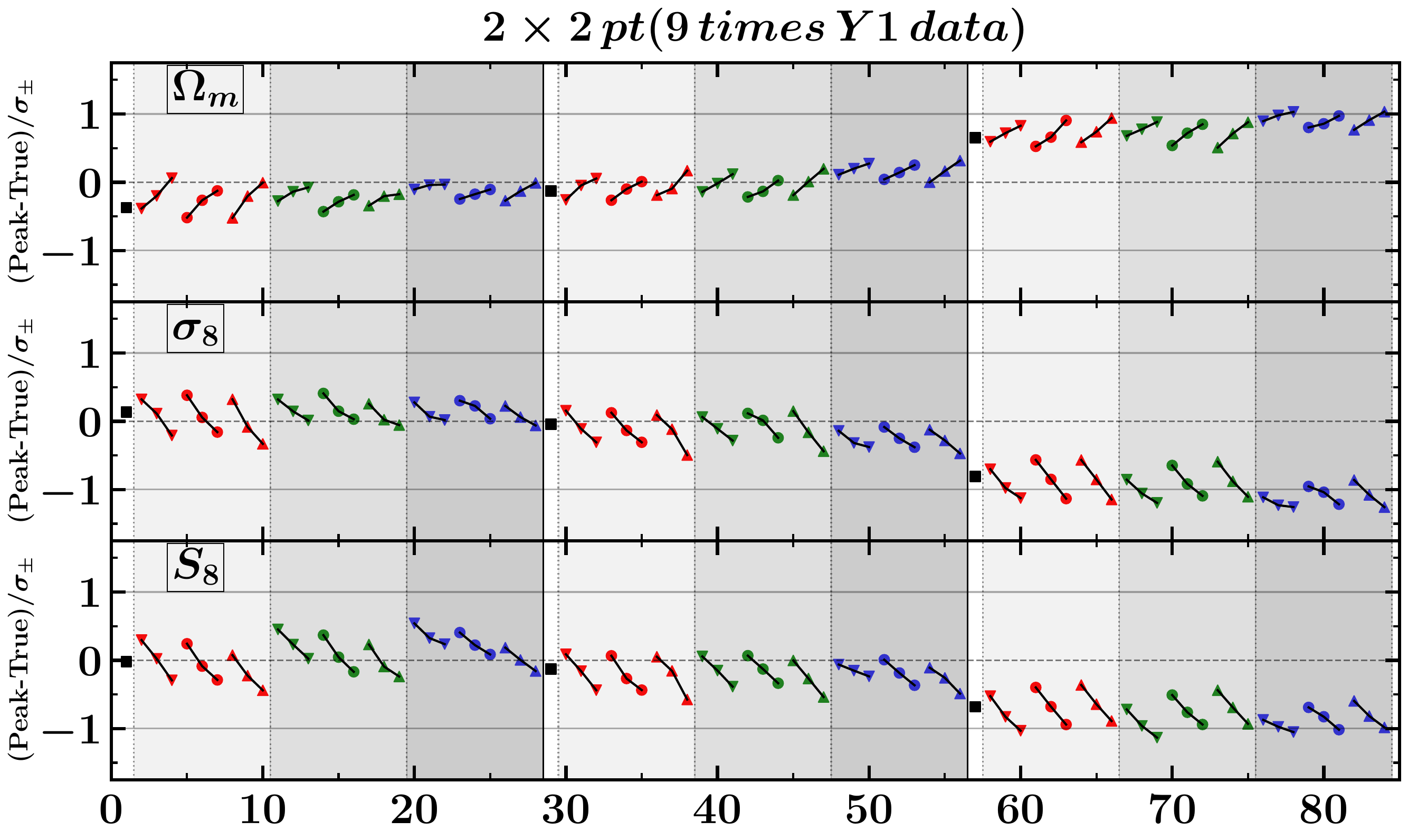}
\includegraphics*[width=0.48\textwidth]{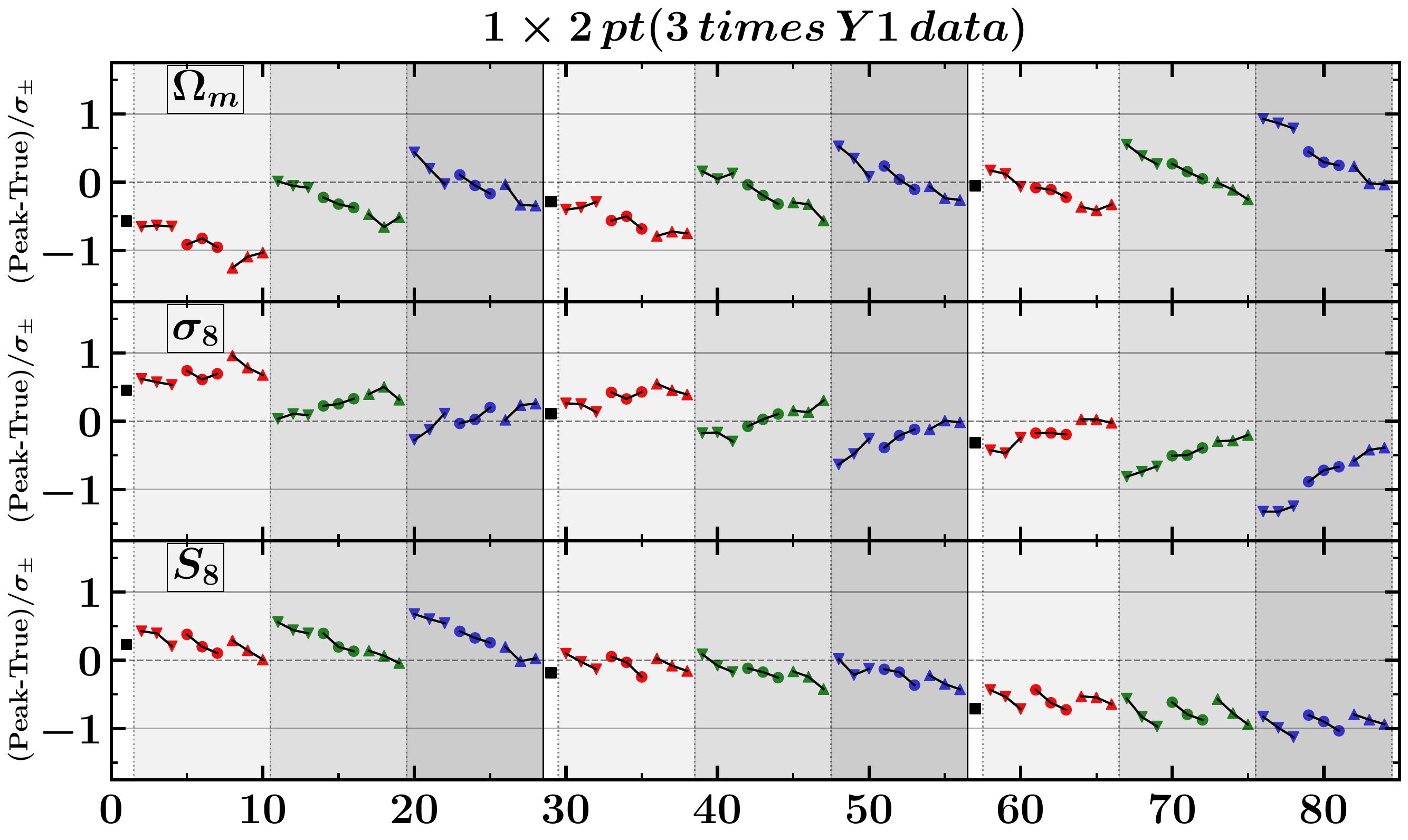}
\includegraphics*[width=0.48\textwidth]{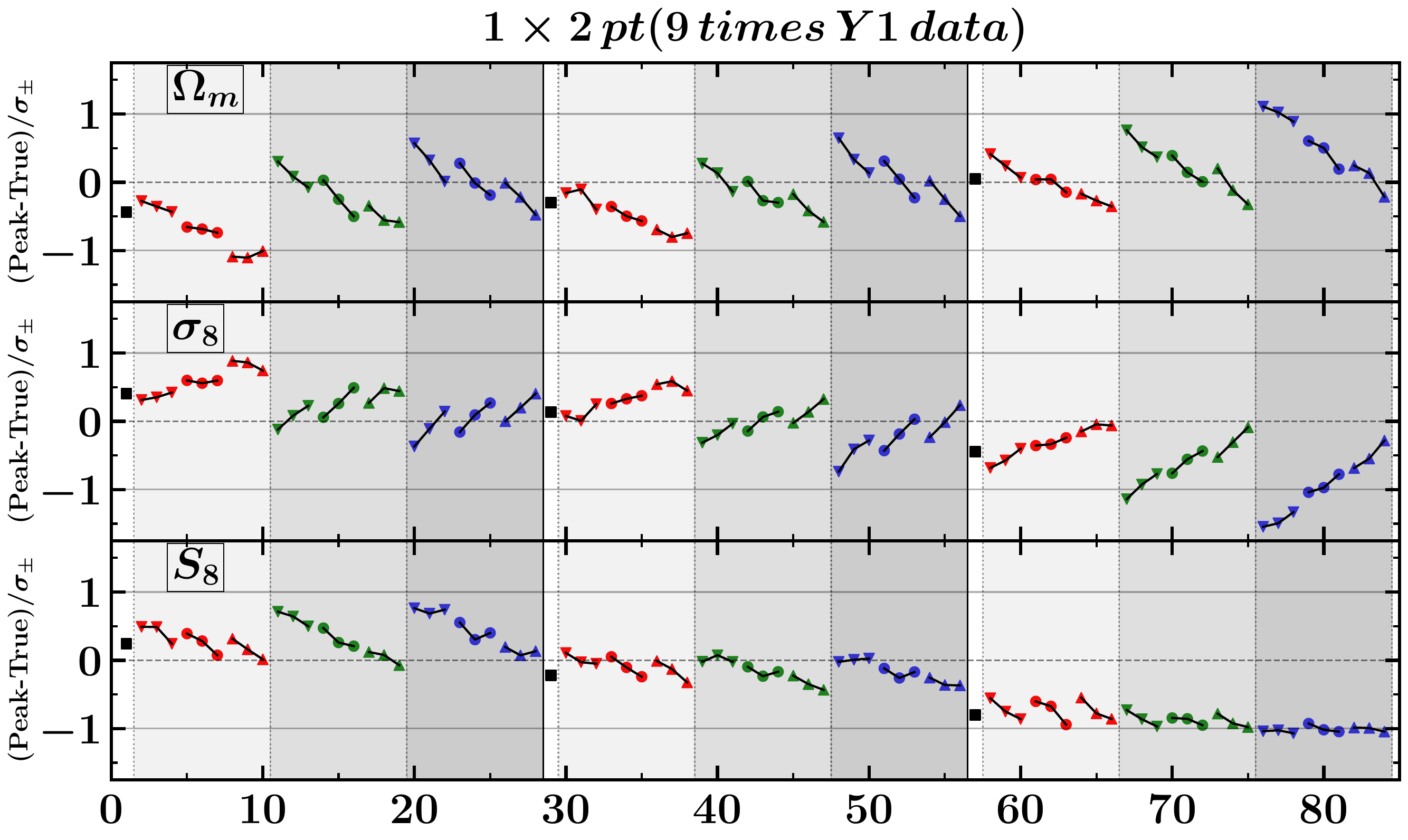}
\caption{These plots show the projection biases in the peaks of the $\Omega_{m}$, $\sigma_{8}$ and $S_{8}$ posteriors for a simulated increase in statistics.  The covariance matrix was reduced by a factor of three in two left plots, and nine in the two right plots.  The top (bottom) two plots correspond to the 2x2pt (1x2pt) analysis. The plots follow the same convention of the plots shown in Figure \ref{fig:peak_bias}.  The factors of three and nine were selected to simulate the recently released DES Y3, and the future DES Y6, Weak Lensing analysis.  See text for more details.}
\label{fig:peak_bias_all}
\end{figure*}
As experiments, like DES, keep increasing the data samples available for WL analysis the likelihoods will get narrower and the prior intervals, if left unchanged, will eventually become much larger than the likelihood widths.  At this point, the projection biases studied in this paper could disappear.  The question, of course, is how much data is needed for this to happen.  In order to answer this question, in this section we study the biases in the peaks of the posteriors simulating an increase in the data size of a factor of three and nine times the sample size used in the DES Y1 WL analysis.  The factor of three was selected because that is the increase in the amount of data going from the DES Y1 to Y3 WL analysis.  The full DES data sample, DES Y6, contains a factor of six times more data than the Y1 sample but we optimistically assumed a factor of nine increase in the amount of data to account for possible improvements in the pipeline and analysis methods.  
The results of the analysis are shown in Figure \ref{fig:peak_bias_all}. The vertical axis in these plots shows the relative bias in the peaks of the posteriors as defined in Equation \ref{eq:pull_definition} and also displayed in Figure \ref{fig:peak_bias}.  The horizontal axis displays the 84 cases listed in Table \ref{tab:variables}, using the same nomenclature as used in Figure \ref{fig:peak_bias}.  The top two plots correspond to the 2x2pt analysis and the bottom two plots to the 1x2pt analysis.  In all cases the analysis was performed on the synthetic data vectors described in Section \ref{subsec:peaks} but, in the two plots on the left, the Y1 covariance matrix was divided by a factor of three, and in the two right plots by a factor of nine.

The results in Figure \ref{fig:peak_bias_all}, when compared with those in Figure \ref{fig:peak_bias}, show that for the 1x2pt analysis the potential increase in constraining power due to the increase in statistics is not enough to reduce the projection bias in the peaks of the posteriors for the parameters $\Omega_{m}$, $\sigma_{8}$, and $S_8$ which continue to reach one $\sigma$ biases for low neutrino masses. The biases in $S_{8}$ for the 1x2pt analysis, even though slightly reduced as the statistical power increases, are still significant in the cases where the true value of $\Omega_\nu h^2$ is low and close to the edge or prior interval (right group of 27 points). The results of the 2x2pt analysis show that the increase in statistics could potentially reduce the projection biases in some of the posterior peaks for the parameters $\Omega_{m}$, $\sigma_{8}$ and $S_{8}$ but biases continue to be significant when the true value of $\Omega_\nu h^2$ is close to the lower edge of the prior interval.

We did not perform ensemble tests for these simulated larger data sets and therefore do not have enough information (e.g. $\sigma_p$ from the pull histograms) to estimate the biases in the 68.27\% credible intervals or the $P_{68}$ coverage probabilities.  However the persistence of the projection biases in the peaks of the posteriors, even when assuming a factor of nine increase in the amount of data, warrants a detailed study of the projection biases in both the peaks of the posteriors and the coverage of credible intervals in, for example, the DES Y6 WL analysis.

%%%%%%%%%%%%%%%%%%%%%%%%%%%%%%%%%%%%%%%%%%%%%%%%%%%%%%%%%%%%%%%%%%%%%%
%%%%%%%%%%%%%%%%%%%%%%%%%%%% Conclusions  %%%%%%%%%%%%%%%%%%%%%%%%%%%%
%%%%%%%%%%%%%%%%%%%%%%%%%%%%%%%%%%%%%%%%%%%%%%%%%%%%%%%%%%%%%%%%%%%%%%

\section{Conclusions}\label{sec:conclusions}
The priors in any analysis reflect the degree of belief of the analyst before the measurement was performed. The community may not share the same degree of belief about the parameter priors used in an experiment. As stated in section 40.2.6.1 of the Particle Data Group(PDG 2022) \cite{PDG}, `` ... For the [Bayesian] result to be of value to the broader community, whose members may not share these [prior] beliefs, it is important to carry out a sensitivity analysis, that is, to show how the result changes under a reasonable variation of the prior probabilities." The projection bias studies in this paper show how the result changes by varying the relative position of the true value with respect to the prior probabilities. These studies, along with previously published results \cite{PRD_us}, show that projection biases of more than $1\sigma$ exist independent of the type of WL analysis (whether it is a 3x2pt, 2x2pt, or 1x2pt analysis). These biases are the result of marginalizing over the parameters $h$, $\Omega_b$, $n_s$, and $\Omega_\nu h^2$ which are correlated with the measured parameters $\Omega_{m}$, $\sigma_{8}$, and $S_{8}$, and have wide likelihood with respect to their priors for the given data set.  The marginalization with respect to the other parameters listed in Tables \ref{tab:parameters} and \ref{tab:parameters2} produce negligible biases because these parameters are either uncorrelated with $\Omega_{m}$, $\sigma_{8}$, and $S_{8}$ or, as in the case of nuisance parameters, the priors are so narrow that they effectively fix the values of these parameters \cite{PRD_us}.  
The results of the study also show that the credible intervals are consistently inflated even in cases where there are no biases in the posterior peaks. Due to this error inflation, and the biases in the posterior peaks, the coverage probability of a $68.27\%$ credible interval could be as large as $93\%$ or as low as $16\%$.
This lack of coverage indicates that, unless corrected, credible intervals in Weak Lensing analysis should not be referred to as confidence intervals (or confidence regions) as has been done in several DES \cite{DES_Y3_1x2pt-a,DES_Y3_1x2pt-b,DES_Y3_2x2pt-a,DES_Y3_2x2pt-b,DES_Y3_3x2pt} and HSC \cite{HSC_2020} publications.  The more appropriate term ``credible intervals" should be used, as has been done recently in the joint cosmic shear analysis by the Dark Energy Survey and the Kilo-Degree Survey collaborations \cite{KiDS-DES}. For several choices of parameters, the pull plots in our projection bias study provide the information to correct biases and transform the credible intervals into confidence intervals for the DES Y1 WL analysis. These, or similar corrections, can be used to quote both the frequentist confidence interval and the Bayesian credible interval showing a clear separation between the two (as suggested in section 40.4 of PDG 2022 \cite{PDG}).  A recent study \cite{eBoss_study} highlights the importance of combining Bayesian and frequentist approaches for a fully nuanced cosmological inference.

The results presented in this paper also show that, for any Bayesian analysis with a low ratio of prior to likelihood widths, the credible intervals do not scale as $1/\sqrt{n}$ (as usually expected) when the experimental data increases by a factor of $n$. Our studies show that the DES credible intervals do not scale as $1/\sqrt{n}$ even when there is six times more data than in the Year 1 (Y1) data sample. 
This indicates that, even with a large increase in statistics, the selected priors limit the reduction of the width of credible intervals with an increase in the amount of data available in the analysis. 
For six out of nine cases tested in the DES Y1 WL analysis, the reduction in errors, when using frequentist confidence intervals instead of credible intervals, is equivalent to an increase of a factor of three in the amount of data.
This indicates a two fold advantage in also quoting confidence intervals: i) the benefit of providing the community with frequentist errors, and ii) a reduction of error bars equivalent to years of data collection and processing efforts.

%%%%%%%%%%%%%%%%%%%%%%%%% acknowledgments  %%%%%%%%%%%%%%%%%%%%%%%%%%

\begin{acknowledgments}
Prudhvi Chintalapati would like to thank the Visiting Scholar Award Program of the Universities Research Association for providing the funding to visit Fermilab and be an integral part of the studies presented in this paper.  This manuscript has been authored by Fermi Research Alliance, LLC under Contract No. DE-AC02-07CH11359 with the U.S. Department of Energy, Office of High Energy Physics.
\end{acknowledgments}

%%%%%%%%%%%%%%%%%%%%%%%%%%%%%%%%%%%%%%%%%%%%%%%%%%%%%%%%%%%%%%
%%%%%%%%%%%%%%%%%%%%%%%%% Appendix  %%%%%%%%%%%%%%%%%%%%%%%%%%
%%%%%%%%%%%%%%%%%%%%%%%%%%%%%%%%%%%%%%%%%%%%%%%%%%%%%%%%%%%%%%

\appendix
%%%%%%%%%%%%%%%%%%%%%%%%%%%%%%%%%%%%%%%%%%%%%%%%%%%%%%%%%%%%%%%%%%%%%%
%%%%%%%%%%%%%%%%%%%%%%%%%%%  Estimations  %%%%%%%%%%%%%%%%%%%%%%%%%%%%
%%%%%%%%%%%%%%%%%%%%%%%%%%%%%%%%%%%%%%%%%%%%%%%%%%%%%%%%%%%%%%%%%%%%%%
\section{Estimation of coverage probabilities}\label{appendix:estimations}
\begin{figure}[htb!]
\includegraphics[width=0.48\textwidth]{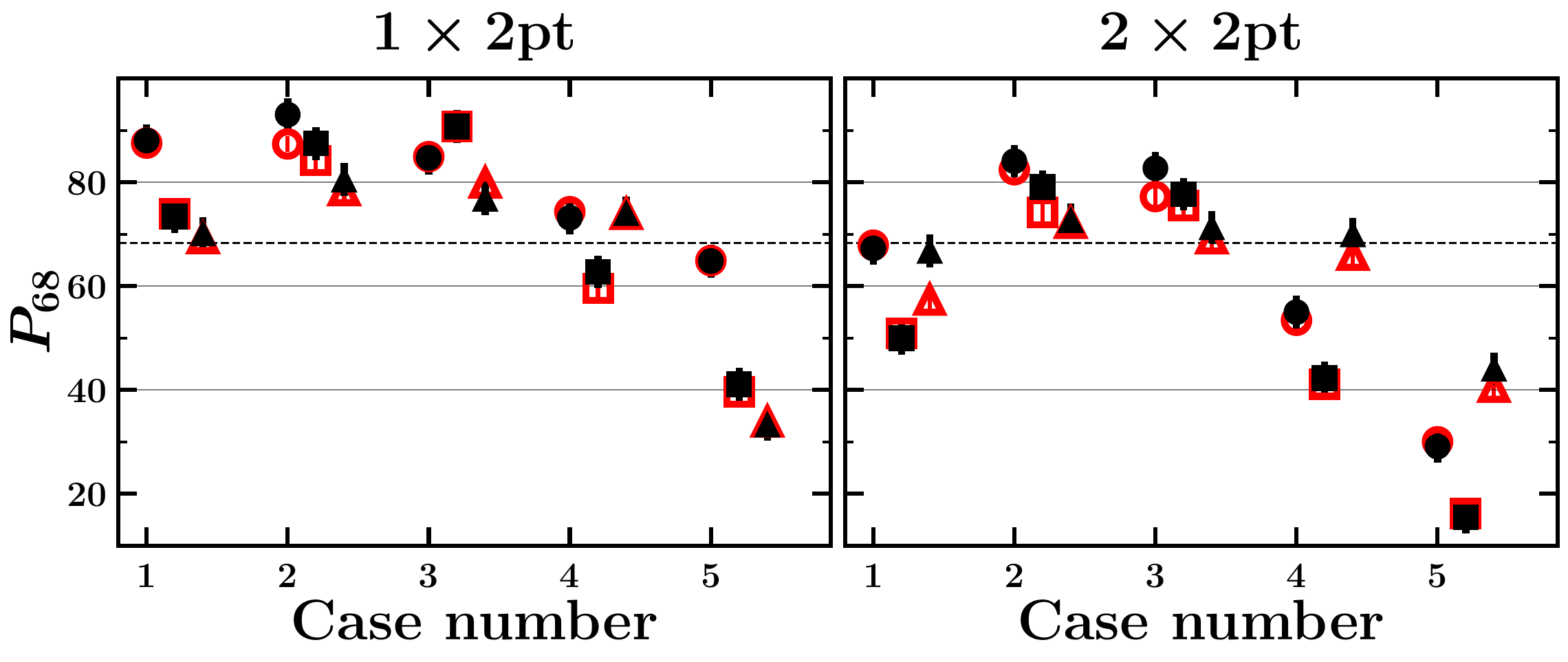}
\caption{The solid markers in this figure show the values of $P_{68}$ calculated counting, in ensemble tests, how often the true value fall inside the credible interval.  These are the $P_{68}$ values shown in Figure \ref{fig:pull_plot}.  The open markers show the values of $P_{68}$ calculated using Equation \ref{eq:p68_pull}.  The left plot corresponds to the 1x2pt WL analysis and the right plot to the 2x2pt one.  The circles, squares, and triangles correspond to the results for $\Omega_{m}$, $\sigma_{8}$, and $S_{8}$ respectively.  The five groups of three points correspond to the five cases listed in Table \ref{tab:ensembles}. The dashed line is the nominal 68.27\% value for $P_{68}$.}
\label{fig:p68_check}
\end{figure}

In Section \ref{subsec:intervals} we calculated the coverage probability $P_{68}$ by doing ensemble tests and counting the number of times the true value of a parameter falls within the corresponding credible interval.  Assuming a Gaussian distribution, and using the Gaussian mean $\bar{x}_{p}$ and width $\sigma_{p}$, one can also calculate $P_{68}$ and its error as (see Appendix C in reference \cite{PRD_us}):
\begin{equation}
\begin{aligned}
P_{68}&= \frac{1}{2} \left\{ \mbox{erf}\left(  \frac{1-\bar{x}_p}{\sqrt{2} \; \sigma_p} \right) + \mbox{erf}\left(  \frac{1+\bar{x}_p}{\sqrt{2} \; \sigma_p} \right)  \right\} \\
\sigma_{P_{68}} &= \sqrt{ \left( \frac{\partial P_{68}}{\partial \bar{x}_p} \, \Delta \bar{x}_p  \right)^2 + \left( \frac{\partial P_{68}}{\partial \sigma_p} \, \Delta \sigma_p  \right)^2 }
\end{aligned}
\label{eq:p68_pull}
\end{equation}
Then, after doing ensemble tests, we can calculate $P_{68}$ by counting or by using Equation \ref{eq:p68_pull} and the values of $\bar{x}_{p}$ and $\sigma_{p}$ obtained from fitting a Gaussian to the pull histograms.  If the pull histogram is a perfect Gaussian then both calculations should give the same answer.  
These two ways of calculating $P_{68}$ were applied to the pull histograms in Figure \ref{fig:pull_plot}.  The results are shown in Figure \ref{fig:p68_check} with the solid markers showing the results calculated by counting, and the open markers showing the results calculated using Equation \ref{eq:p68_pull}. We can see that both ways of calculating the coverage probability gives, within errors, the same results.

\begin{figure}[hbt]
\includegraphics[width=0.48\textwidth]{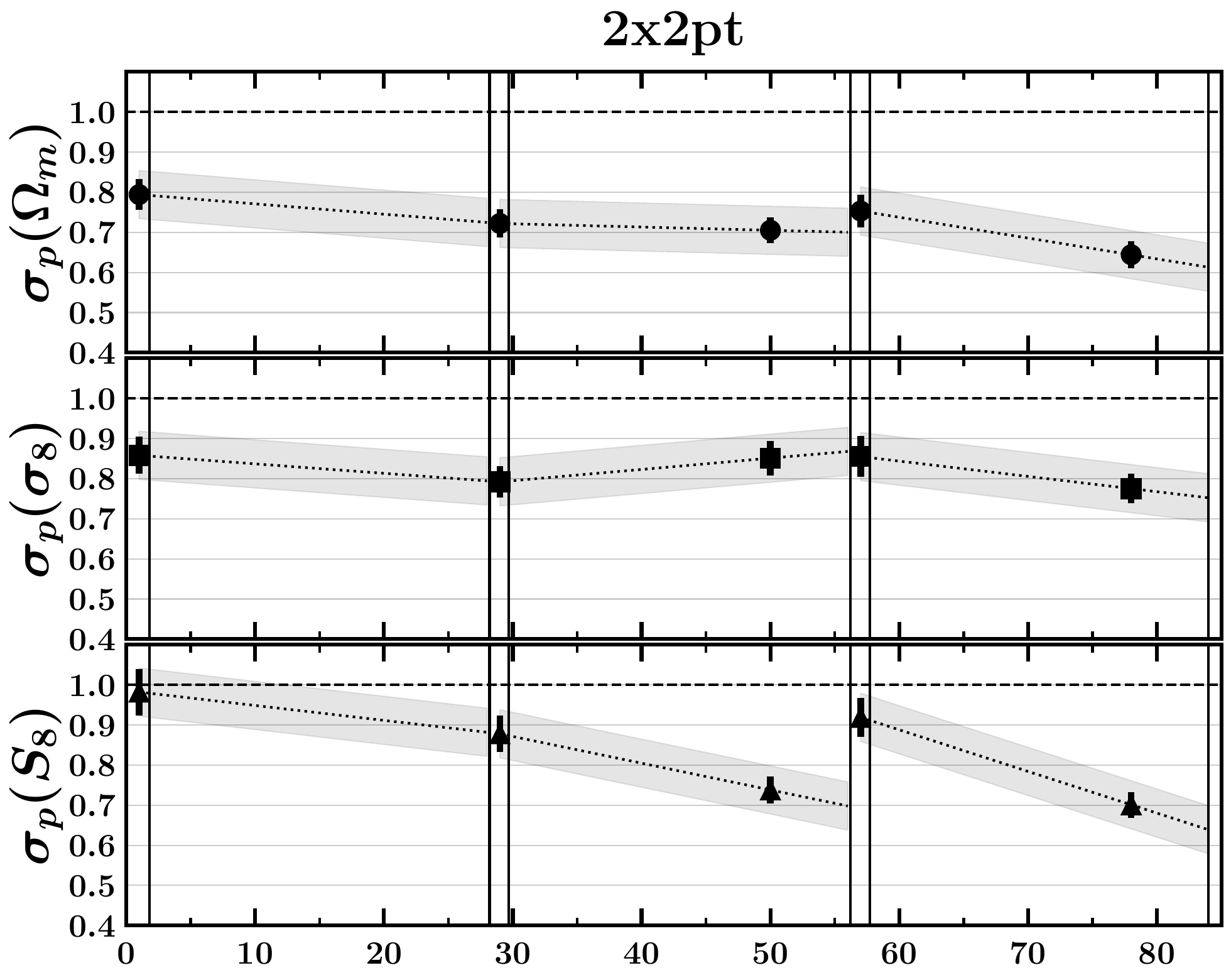}
\includegraphics[width=0.48\textwidth]{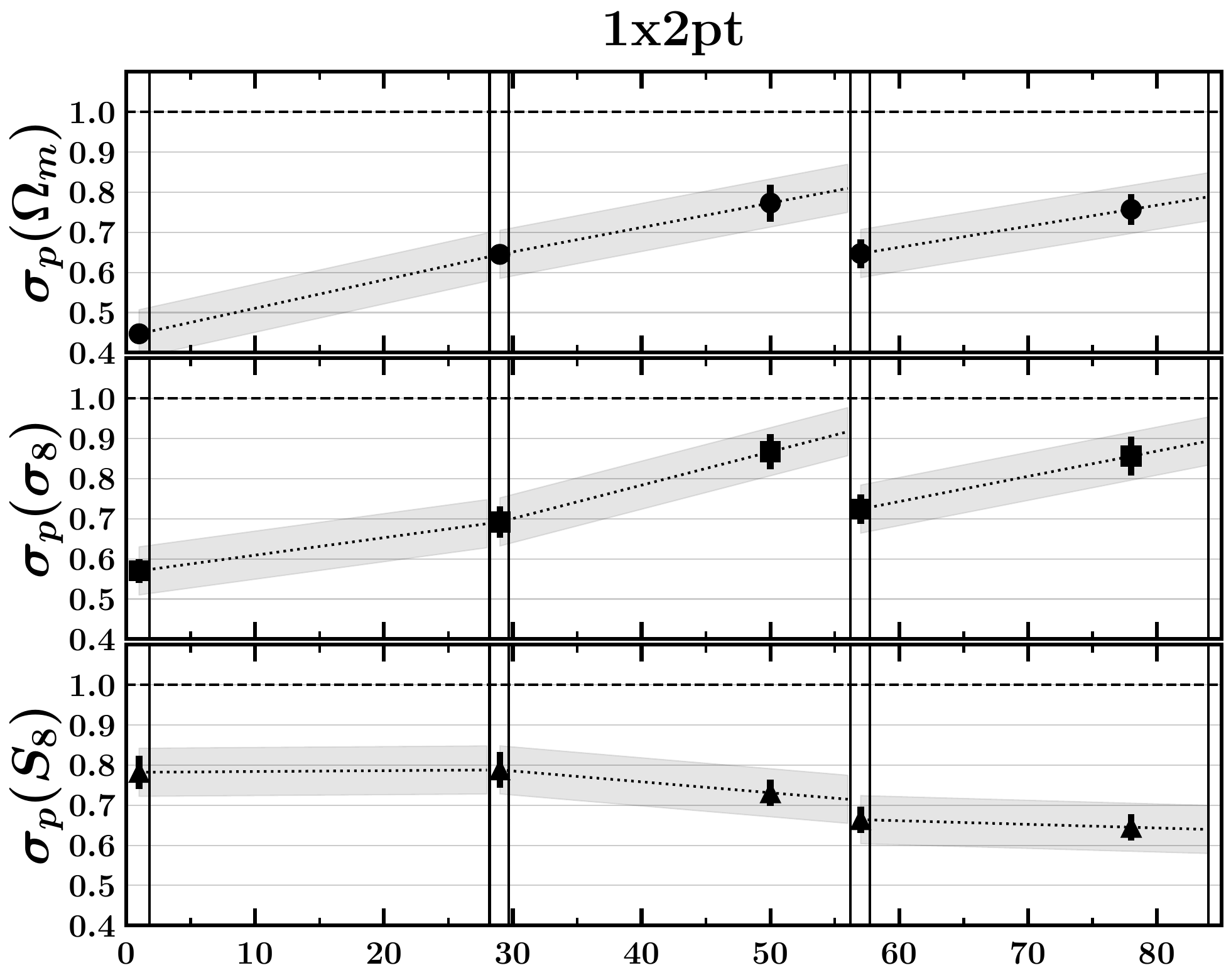}
\caption{The black markers with error bars in the plots, show the width $\sigma_p$ obtained from the Gaussian fits in Figure \ref{fig:pull_plot}.  The circles, squares and triangles represent the results for the parameters $\Omega_{m}$, $\sigma_{8}$ and $S_{8}$ respectively. The horizontal axis is the same as in Figure \ref{fig:peak_bias}. The dotted lines with the grey error bands show the values of $\sigma_p$ used to estimate the $P_{68}$ probabilities shown in Figure \ref{fig:P68_all}.}
\label{fig:pull_sigma_bands}
\end{figure}

\begin{figure}[htb]
\includegraphics[width=0.48\textwidth]{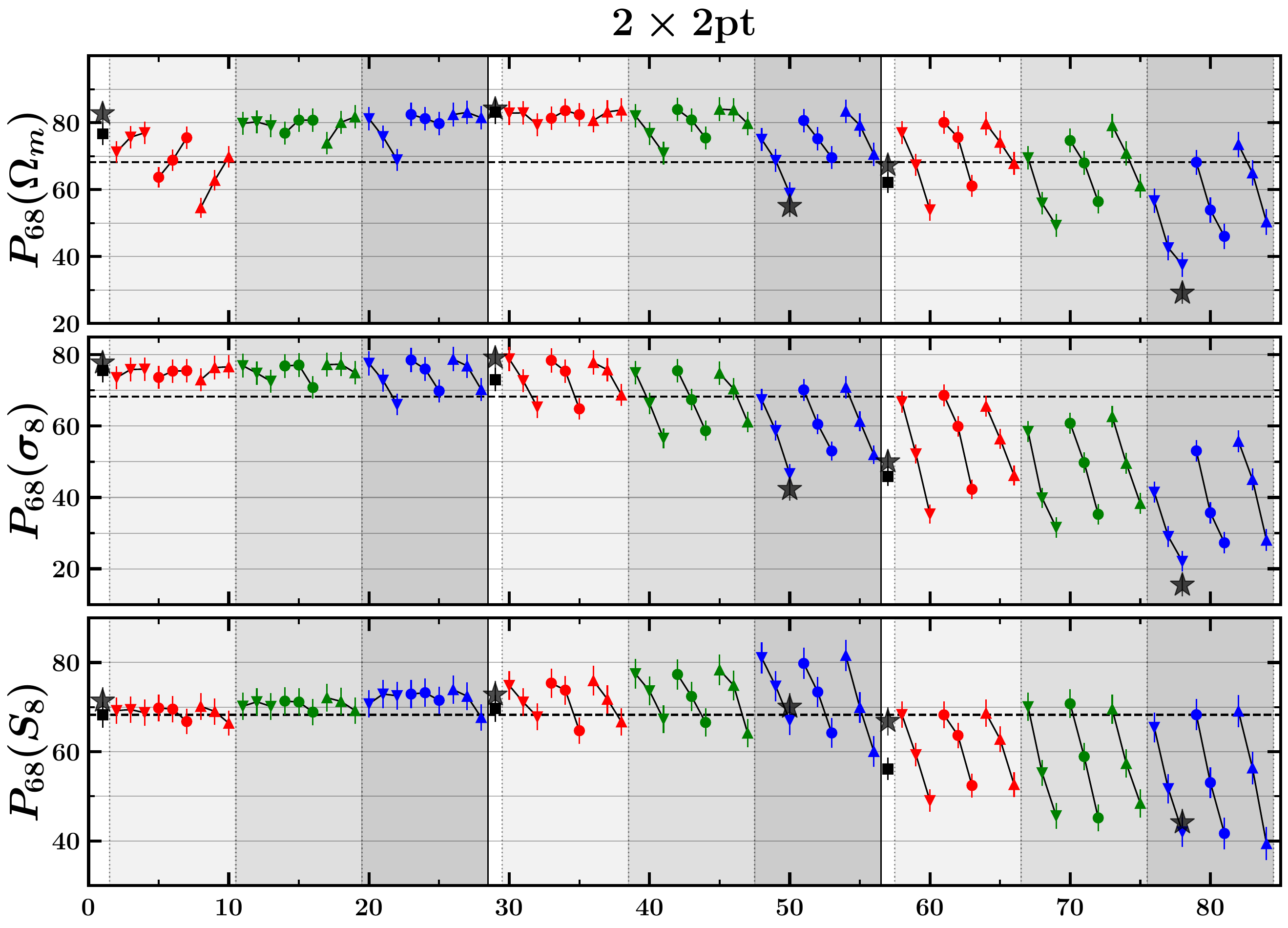}
\includegraphics[width=0.48\textwidth]{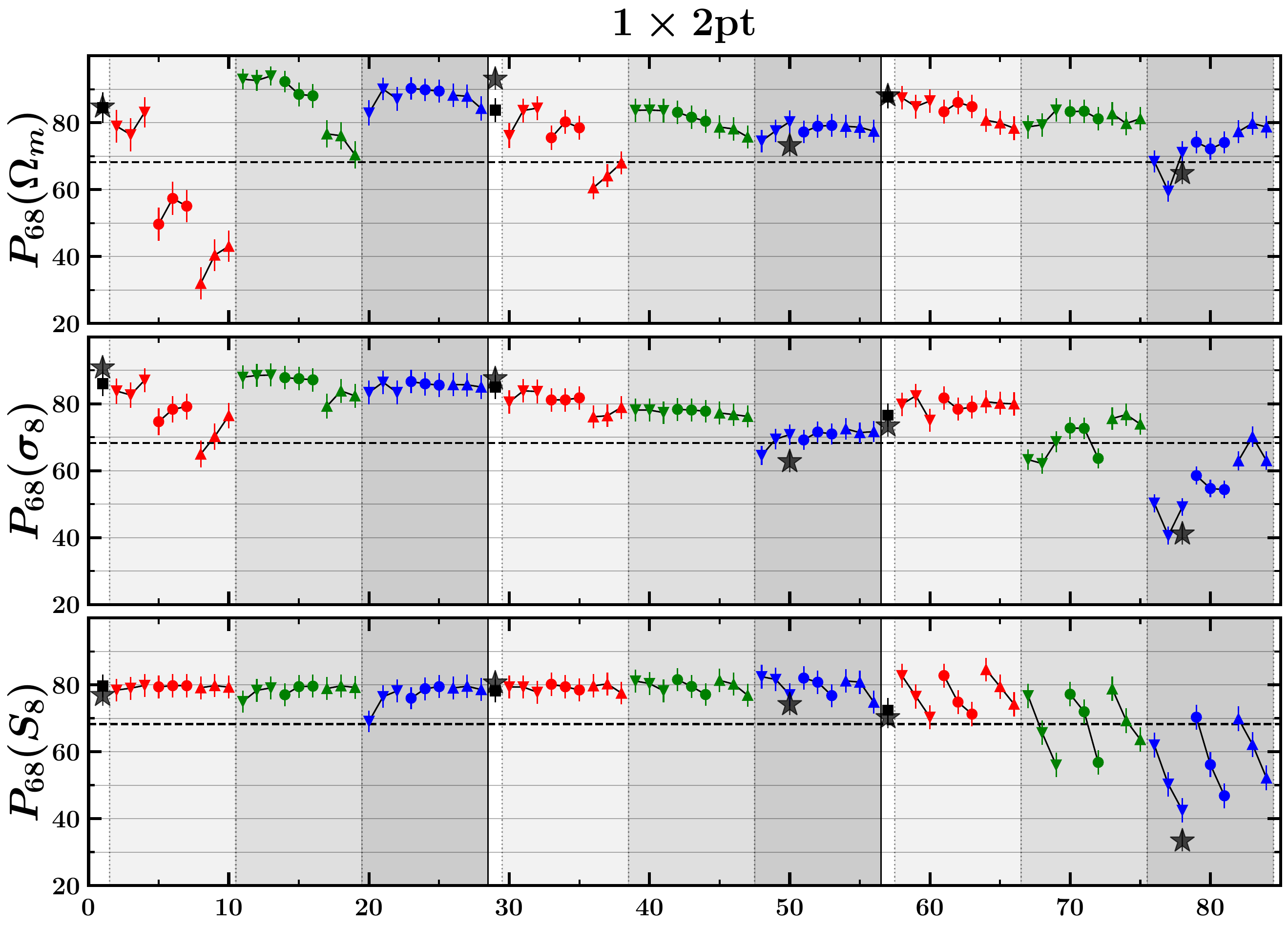}
\caption{Best estimate of the $P_{68}$ probability for the 84 cases shown in Figure \ref{fig:peak_bias}.  $P_{68}$ is the coverage probability of the $68.27\%$ credible interval. See text for details of the $P_{68}$ calculation.  The 2x2pt (1x2pt) WL analysis results are shown in the top (bottom) plot.  The vertical axes show $P_{68}$, in percent, for $\Omega_m$, $\sigma_8$, and $S_8$.  The horizontal axis, marker styles, and shades in the plot follow Figure \ref{fig:peak_bias}.  The five solid black stars with error bars show the values of $P_{68}$ calculated by counting as in Figure \ref{fig:pull_plot}.}
\label{fig:P68_all}
\end{figure}

As we just described, we can get a good measurement of the coverage probability only for the five cases, out of 84, for which we performed ensemble tests.  For the rest, we can estimate the values of $P_{68}$ by making two reasonable assumptions.  As shown in Figure \ref{fig:posteriors-compare}, the value of $\bar{x}_p$ obtained from the pull plots is very similar to the peak of the posterior calculated in the analysis of the synthetic data vectors. Therefore, we can use the 84 posterior peaks used to produce Figure \ref{fig:pull_plot} as an approximation for $\bar{x}_p$. For $\sigma_p$, we will interpolate between the values calculated from the ensemble tests.  This interpolation is shown in Figure \ref{fig:pull_sigma_bands}. Based on the behaviour of the posterior peaks in Figure \ref{fig:peak_bias} we assumed a linear dependence in $\sigma_p$ for each of the $\Omega_{\nu} h^{2}$ regions.  We assigned an error band to this interpolation of $0.06$, which is larger than the largest error bars observed in all the pull plots.  The results of $P_{68}$, calculated using these two approximations and Equation \ref{eq:p68_pull}, are shown in Figure \ref{fig:P68_all}.


The results shown in Figure \ref{fig:P68_all} for the 2x2pt WL analysis are very similar to the 3x2pt WL analysis coverage probabilities shown in Figure 9 of reference \cite{PRD_us}.  In a large number of cases, $P_{68}$ is larger than 68.27\%, which is not surprising given the fact that $\sigma_p$ is systematically less than one. However, for low values of $\Omega_{\nu} h^{2}$, the biases in the posterior peaks are large enough that, even with inflated errors, the coverage probabilities can be as low as 20\%.  For the 1x2pt analysis, shown in the bottom plot of Figure \ref{fig:P68_all}, the coverage probability is more uniform than in the 2x2pt analysis, but in the majority of cases is still far from the expected value of 68.27\%.  

\bibliography{PRD_Main}% Produces the bibliography via BibTeX.

\providecommand{\noopsort}[1]{}\providecommand{\singleletter}[1]{#1}%
\begin{thebibliography}{32}%
\makeatletter
\providecommand \@ifxundefined [1]{%
 \@ifx{#1\undefined}
}%
\providecommand \@ifnum [1]{%
 \ifnum #1\expandafter \@firstoftwo
 \else \expandafter \@secondoftwo
 \fi
}%
\providecommand \@ifx [1]{%
 \ifx #1\expandafter \@firstoftwo
 \else \expandafter \@secondoftwo
 \fi
}%
\providecommand \natexlab [1]{#1}%
\providecommand \enquote  [1]{``#1''}%
\providecommand \bibnamefont  [1]{#1}%
\providecommand \bibfnamefont [1]{#1}%
\providecommand \citenamefont [1]{#1}%
\providecommand \href@noop [0]{\@secondoftwo}%
\providecommand \href [0]{\begingroup \@sanitize@url \@href}%
\providecommand \@href[1]{\@@startlink{#1}\@@href}%
\providecommand \@@href[1]{\endgroup#1\@@endlink}%
\providecommand \@sanitize@url [0]{\catcode `\\12\catcode `\$12\catcode
  `\&12\catcode `\#12\catcode `\^12\catcode `\_12\catcode `\%12\relax}%
\providecommand \@@startlink[1]{}%
\providecommand \@@endlink[0]{}%
\providecommand \url  [0]{\begingroup\@sanitize@url \@url }%
\providecommand \@url [1]{\endgroup\@href {#1}{\urlprefix }}%
\providecommand \urlprefix  [0]{URL }%
\providecommand \Eprint [0]{\href }%
\providecommand \doibase [0]{https://doi.org/}%
\providecommand \selectlanguage [0]{\@gobble}%
\providecommand \bibinfo  [0]{\@secondoftwo}%
\providecommand \bibfield  [0]{\@secondoftwo}%
\providecommand \translation [1]{[#1]}%
\providecommand \BibitemOpen [0]{}%
\providecommand \bibitemStop [0]{}%
\providecommand \bibitemNoStop [0]{.\EOS\space}%
\providecommand \EOS [0]{\spacefactor3000\relax}%
\providecommand \BibitemShut  [1]{\csname bibitem#1\endcsname}%
\let\auto@bib@innerbib\@empty
%</preamble>
\bibitem [{\citenamefont {Fisher}\ \emph {et~al.}(2000)\citenamefont {Fisher}
  \emph {et~al.}}]{Sloan_WL}%
  \BibitemOpen
  \bibfield  {author} {\bibinfo {author} {\bibfnamefont {P.}~\bibnamefont
  {Fisher}} \emph {et~al.} (\bibinfo {collaboration} {SDSS}),\ }\href@noop {}
  {\bibfield  {journal} {\bibinfo  {journal} {The Astronomical Journal}\
  }\textbf {\bibinfo {volume} {120}},\ \bibinfo {pages} {1198} (\bibinfo {year}
  {2000})}\BibitemShut {NoStop}%
\bibitem [{\citenamefont {Wittman}\ \emph {et~al.}(2000)\citenamefont
  {Wittman}, \citenamefont {Tyson}, \citenamefont {Kirkman}, \citenamefont
  {Dell’Antonio},\ and\ \citenamefont {Bernstein}}]{Nature_2000_WL}%
  \BibitemOpen
  \bibfield  {author} {\bibinfo {author} {\bibfnamefont {D.~M.}\ \bibnamefont
  {Wittman}}, \bibinfo {author} {\bibfnamefont {J.~A.}\ \bibnamefont {Tyson}},
  \bibinfo {author} {\bibfnamefont {D.}~\bibnamefont {Kirkman}}, \bibinfo
  {author} {\bibfnamefont {I.}~\bibnamefont {Dell’Antonio}},\ and\ \bibinfo
  {author} {\bibfnamefont {G.}~\bibnamefont {Bernstein}},\ }\href@noop {}
  {\bibfield  {journal} {\bibinfo  {journal} {Nature}\ }\textbf {\bibinfo
  {volume} {405}},\ \bibinfo {pages} {143} (\bibinfo {year}
  {2000})}\BibitemShut {NoStop}%
\bibitem [{\citenamefont {Bacon}\ \emph {et~al.}(2000)\citenamefont {Bacon},
  \citenamefont {Refregier},\ and\ \citenamefont {Ellis}}]{MNRAS_2000_WL}%
  \BibitemOpen
  \bibfield  {author} {\bibinfo {author} {\bibfnamefont {D.~J.}\ \bibnamefont
  {Bacon}}, \bibinfo {author} {\bibfnamefont {A.~R.}\ \bibnamefont
  {Refregier}},\ and\ \bibinfo {author} {\bibfnamefont {R.~S.}\ \bibnamefont
  {Ellis}},\ }\href@noop {} {\bibfield  {journal} {\bibinfo  {journal} {MNRAS}\
  }\textbf {\bibinfo {volume} {318}},\ \bibinfo {pages} {625} (\bibinfo {year}
  {2000})}\BibitemShut {NoStop}%
\bibitem [{\citenamefont {van Waerbeke}\ \emph {et~al.}(2000)\citenamefont {van
  Waerbeke} \emph {et~al.}}]{A&A_2000_WL}%
  \BibitemOpen
  \bibfield  {author} {\bibinfo {author} {\bibfnamefont {L.}~\bibnamefont {van
  Waerbeke}} \emph {et~al.} (\bibinfo {collaboration} {CFHT}),\ }\href@noop {}
  {\bibfield  {journal} {\bibinfo  {journal} {A\&A}\ }\textbf {\bibinfo
  {volume} {358}},\ \bibinfo {pages} {30} (\bibinfo {year} {2000})}\BibitemShut
  {NoStop}%
\bibitem [{DES()}]{DES_WEB}%
  \BibitemOpen
  \href@noop {} {\bibinfo {title} {Dark energy survery (des)
  \url{https://www.darkenergysurvey.org/}.}}\BibitemShut {Stop}%
\bibitem [{\citenamefont {Amon}\ \emph {et~al.}(2022)\citenamefont {Amon} \emph
  {et~al.}}]{DES_Y3_1x2pt-a}%
  \BibitemOpen
  \bibfield  {author} {\bibinfo {author} {\bibfnamefont {A.}~\bibnamefont
  {Amon}} \emph {et~al.} (\bibinfo {collaboration} {DES}),\ }\href
  {https://doi.org/10.1103/PhysRevD.105.023514} {\bibfield  {journal} {\bibinfo
   {journal} {Phys. Rev. D}\ }\textbf {\bibinfo {volume} {105}},\ \bibinfo
  {pages} {023514} (\bibinfo {year} {2022})}\BibitemShut {NoStop}%
\bibitem [{\citenamefont {Secco}\ \emph {et~al.}(2022)\citenamefont {Secco}
  \emph {et~al.}}]{DES_Y3_1x2pt-b}%
  \BibitemOpen
  \bibfield  {author} {\bibinfo {author} {\bibfnamefont {L.~F.}\ \bibnamefont
  {Secco}} \emph {et~al.} (\bibinfo {collaboration} {DES}),\ }\href
  {https://doi.org/10.1103/PhysRevD.105.023515} {\bibfield  {journal} {\bibinfo
   {journal} {Phys. Rev. D}\ }\textbf {\bibinfo {volume} {105}},\ \bibinfo
  {pages} {023515} (\bibinfo {year} {2022})}\BibitemShut {NoStop}%
\bibitem [{\citenamefont {Pandey}\ \emph {et~al.}(2022)\citenamefont {Pandey}
  \emph {et~al.}}]{DES_Y3_2x2pt-a}%
  \BibitemOpen
  \bibfield  {author} {\bibinfo {author} {\bibfnamefont {S.}~\bibnamefont
  {Pandey}} \emph {et~al.} (\bibinfo {collaboration} {DES}),\ }\href
  {https://doi.org/10.1103/PhysRevD.106.043520} {\bibfield  {journal} {\bibinfo
   {journal} {Phys. Rev. D}\ }\textbf {\bibinfo {volume} {106}},\ \bibinfo
  {pages} {043520} (\bibinfo {year} {2022})}\BibitemShut {NoStop}%
\bibitem [{\citenamefont {Porredon}\ \emph {et~al.}(2021)\citenamefont
  {Porredon} \emph {et~al.}}]{DES_Y3_2x2pt-b}%
  \BibitemOpen
  \bibfield  {author} {\bibinfo {author} {\bibfnamefont {A.}~\bibnamefont
  {Porredon}} \emph {et~al.} (\bibinfo {collaboration} {DES}),\ }\href@noop {}
  {\bibfield  {journal} {\bibinfo  {journal} {arXiv:2105.13546}\ } (\bibinfo
  {year} {2021})}\BibitemShut {NoStop}%
\bibitem [{Euc()}]{Euclid_WEB}%
  \BibitemOpen
  \href@noop {} {\bibinfo {title} {Euclid
  \url{https://www.euclid-ec.org}.}}\BibitemShut {Stop}%
\bibitem [{LSS()}]{LSST_WEB}%
  \BibitemOpen
  \href@noop {} {\bibinfo {title} {Lsst
  \url{https://www.lsst.org}.}}\BibitemShut {Stop}%
\bibitem [{NRS()}]{NRST_WEB}%
  \BibitemOpen
  \href@noop {} {\bibinfo {title} {Nancy roman space telescope
  \url{https://roman.gsfc.nasa.gov}.}}\BibitemShut {Stop}%
\bibitem [{\citenamefont {Chintalapati}\ \emph {et~al.}(2022)\citenamefont
  {Chintalapati}, \citenamefont {Gutierrez},\ and\ \citenamefont
  {Wang}}]{PRD_us}%
  \BibitemOpen
  \bibfield  {author} {\bibinfo {author} {\bibfnamefont {P.~R.~V.}\
  \bibnamefont {Chintalapati}}, \bibinfo {author} {\bibfnamefont
  {G.}~\bibnamefont {Gutierrez}},\ and\ \bibinfo {author} {\bibfnamefont {M.~H.
  L.~S.}\ \bibnamefont {Wang}},\ }\bibfield  {title} {\bibinfo {title}
  {Systematic study of projection biases in weak lensing analysis},\ }\href
  {https://doi.org/10.1103/PhysRevD.105.043515} {\bibfield  {journal} {\bibinfo
   {journal} {Phys. Rev. D}\ }\textbf {\bibinfo {volume} {105}},\ \bibinfo
  {pages} {043515} (\bibinfo {year} {2022})}\BibitemShut {NoStop}%
\bibitem [{\citenamefont {Abbott}\ \emph {et~al.}(2018)\citenamefont {Abbott}
  \emph {et~al.}}]{DES_Y1_PRD}%
  \BibitemOpen
  \bibfield  {author} {\bibinfo {author} {\bibfnamefont {T.~M.~C.}\
  \bibnamefont {Abbott}} \emph {et~al.} (\bibinfo {collaboration} {DES}),\
  }\href@noop {} {\bibfield  {journal} {\bibinfo  {journal} {Phys.\ Rev. D}\
  }\textbf {\bibinfo {volume} {98}},\ \bibinfo {pages} {043526} (\bibinfo
  {year} {2018})}\BibitemShut {NoStop}%
\bibitem [{\citenamefont {Zuntz}\ \emph {et~al.}(2015)\citenamefont {Zuntz},
  \citenamefont {Paterno}, \citenamefont {Jennings}, \citenamefont {Rudd},
  \citenamefont {Manzotti}, \citenamefont {Dodelson}, \citenamefont {Bridle},
  \citenamefont {Sehrish},\ and\ \citenamefont {Kowalkowski}}]{Cosmosis_paper}%
  \BibitemOpen
  \bibfield  {author} {\bibinfo {author} {\bibfnamefont {J.}~\bibnamefont
  {Zuntz}}, \bibinfo {author} {\bibfnamefont {M.}~\bibnamefont {Paterno}},
  \bibinfo {author} {\bibfnamefont {E.}~\bibnamefont {Jennings}}, \bibinfo
  {author} {\bibfnamefont {D.}~\bibnamefont {Rudd}}, \bibinfo {author}
  {\bibfnamefont {A.}~\bibnamefont {Manzotti}}, \bibinfo {author}
  {\bibfnamefont {S.}~\bibnamefont {Dodelson}}, \bibinfo {author}
  {\bibfnamefont {S.}~\bibnamefont {Bridle}}, \bibinfo {author} {\bibfnamefont
  {S.}~\bibnamefont {Sehrish}},\ and\ \bibinfo {author} {\bibfnamefont
  {J.}~\bibnamefont {Kowalkowski}},\ }\href@noop {} {\bibfield  {journal}
  {\bibinfo  {journal} {Astronomy and Computing}\ }\textbf {\bibinfo {volume}
  {12}},\ \bibinfo {pages} {45} (\bibinfo {year} {2015})}\BibitemShut {NoStop}%
\bibitem [{Cos()}]{Cosmosis_program}%
  \BibitemOpen
  \href@noop {} {\bibinfo {title} {To run cosmosis use the installer available
  at https://bitbucket.org/joezuntz/cosmosis/wiki/home.}}\BibitemShut {Stop}%
\bibitem [{\citenamefont {Lewis}\ \emph {et~al.}(2000)\citenamefont {Lewis},
  \citenamefont {Challinor},\ and\ \citenamefont {Lasenby}}]{CAMB}%
  \BibitemOpen
  \bibfield  {author} {\bibinfo {author} {\bibfnamefont {A.}~\bibnamefont
  {Lewis}}, \bibinfo {author} {\bibfnamefont {A.}~\bibnamefont {Challinor}},\
  and\ \bibinfo {author} {\bibfnamefont {A.}~\bibnamefont {Lasenby}},\
  }\href@noop {} {\bibfield  {journal} {\bibinfo  {journal} {The Astrophysical
  Journal}\ }\textbf {\bibinfo {volume} {538}},\ \bibinfo {pages} {473}
  (\bibinfo {year} {2000})}\BibitemShut {NoStop}%
\bibitem [{\citenamefont {Smith}\ \emph {et~al.}(2003)\citenamefont {Smith},
  \citenamefont {Peacock}, \citenamefont {A.~Jenkins}, \citenamefont {Frenk},
  \citenamefont {Pearce}, \citenamefont {Thomas}, \citenamefont {Efstathiou},\
  and\ \citenamefont {Couchman}}]{Halofit_1}%
  \BibitemOpen
  \bibfield  {author} {\bibinfo {author} {\bibfnamefont {R.~E.}\ \bibnamefont
  {Smith}}, \bibinfo {author} {\bibfnamefont {J.~A.}\ \bibnamefont {Peacock}},
  \bibinfo {author} {\bibfnamefont {S.~D. M.~W.}\ \bibnamefont {A.~Jenkins}},
  \bibinfo {author} {\bibfnamefont {C.~S.}\ \bibnamefont {Frenk}}, \bibinfo
  {author} {\bibfnamefont {F.~R.}\ \bibnamefont {Pearce}}, \bibinfo {author}
  {\bibfnamefont {P.~A.}\ \bibnamefont {Thomas}}, \bibinfo {author}
  {\bibfnamefont {G.}~\bibnamefont {Efstathiou}},\ and\ \bibinfo {author}
  {\bibfnamefont {H.~M.~P.}\ \bibnamefont {Couchman}},\ }\href@noop {}
  {\bibfield  {journal} {\bibinfo  {journal} {Mon. Not. R. Astron. Soc.}\
  }\textbf {\bibinfo {volume} {341}},\ \bibinfo {pages} {1311} (\bibinfo {year}
  {2003})}\BibitemShut {NoStop}%
\bibitem [{\citenamefont {Bird}\ \emph {et~al.}(2012)\citenamefont {Bird},
  \citenamefont {Viel},\ and\ \citenamefont {Haehnelt}}]{Halofit_2}%
  \BibitemOpen
  \bibfield  {author} {\bibinfo {author} {\bibfnamefont {S.}~\bibnamefont
  {Bird}}, \bibinfo {author} {\bibfnamefont {M.}~\bibnamefont {Viel}},\ and\
  \bibinfo {author} {\bibfnamefont {M.~G.}\ \bibnamefont {Haehnelt}},\
  }\href@noop {} {\bibfield  {journal} {\bibinfo  {journal} {Mon. Not. R.
  Astron. Soc.}\ }\textbf {\bibinfo {volume} {420}},\ \bibinfo {pages} {2551}
  (\bibinfo {year} {2012})}\BibitemShut {NoStop}%
\bibitem [{\citenamefont {Takahashi}\ \emph {et~al.}(2012)\citenamefont
  {Takahashi}, \citenamefont {Sato}, \citenamefont {Nishimichi}, \citenamefont
  {Taruya},\ and\ \citenamefont {Oguri}}]{Halofit_3}%
  \BibitemOpen
  \bibfield  {author} {\bibinfo {author} {\bibfnamefont {R.}~\bibnamefont
  {Takahashi}}, \bibinfo {author} {\bibfnamefont {M.}~\bibnamefont {Sato}},
  \bibinfo {author} {\bibfnamefont {T.}~\bibnamefont {Nishimichi}}, \bibinfo
  {author} {\bibfnamefont {A.}~\bibnamefont {Taruya}},\ and\ \bibinfo {author}
  {\bibfnamefont {M.}~\bibnamefont {Oguri}},\ }\href@noop {} {\bibfield
  {journal} {\bibinfo  {journal} {The Astrophysical Journal}\ }\textbf
  {\bibinfo {volume} {761}},\ \bibinfo {pages} {152} (\bibinfo {year}
  {2012})}\BibitemShut {NoStop}%
\bibitem [{\citenamefont {Hirata}\ and\ \citenamefont
  {Seljak}(2004)}]{Hirata-Seljak}%
  \BibitemOpen
  \bibfield  {author} {\bibinfo {author} {\bibfnamefont {C.~M.}\ \bibnamefont
  {Hirata}}\ and\ \bibinfo {author} {\bibfnamefont {U.}~\bibnamefont
  {Seljak}},\ }\href@noop {} {\bibfield  {journal} {\bibinfo  {journal} {Phys.
  Rev. D}\ }\textbf {\bibinfo {volume} {70}},\ \bibinfo {pages} {063526}
  (\bibinfo {year} {2004})}\BibitemShut {NoStop}%
\bibitem [{\citenamefont {Hirata}\ and\ \citenamefont
  {Seljak}(2010)}]{Hirata-Seljak-erratum}%
  \BibitemOpen
  \bibfield  {author} {\bibinfo {author} {\bibfnamefont {C.~M.}\ \bibnamefont
  {Hirata}}\ and\ \bibinfo {author} {\bibfnamefont {U.}~\bibnamefont
  {Seljak}},\ }\href@noop {} {\bibfield  {journal} {\bibinfo  {journal} {Phys.
  Rev. D}\ }\textbf {\bibinfo {volume} {82}},\ \bibinfo {pages} {049901(E)}
  (\bibinfo {year} {2010})}\BibitemShut {NoStop}%
\bibitem [{\citenamefont {Bridle}\ and\ \citenamefont
  {King}(2007)}]{Bridle-King_IA}%
  \BibitemOpen
  \bibfield  {author} {\bibinfo {author} {\bibfnamefont {S.}~\bibnamefont
  {Bridle}}\ and\ \bibinfo {author} {\bibfnamefont {L.}~\bibnamefont {King}},\
  }\href@noop {} {\bibfield  {journal} {\bibinfo  {journal} {New Journal of
  Physics}\ }\textbf {\bibinfo {volume} {9}},\ \bibinfo {pages} {444} (\bibinfo
  {year} {2007})}\BibitemShut {NoStop}%
\bibitem [{\citenamefont {Feroz}\ \emph {et~al.}(2009)\citenamefont {Feroz},
  \citenamefont {Hobson},\ and\ \citenamefont {Bridges}}]{Multinest_ref}%
  \BibitemOpen
  \bibfield  {author} {\bibinfo {author} {\bibfnamefont {F.}~\bibnamefont
  {Feroz}}, \bibinfo {author} {\bibfnamefont {M.~P.}\ \bibnamefont {Hobson}},\
  and\ \bibinfo {author} {\bibfnamefont {M.}~\bibnamefont {Bridges}},\
  }\href@noop {} {\bibfield  {journal} {\bibinfo  {journal} {Mon. Not. R.
  Astron. Soc.}\ }\textbf {\bibinfo {volume} {398}},\ \bibinfo {pages} {1601}
  (\bibinfo {year} {2009})}\BibitemShut {NoStop}%
\bibitem [{\citenamefont {{Hinton}}(2016)}]{Chainconsumer_ref}%
  \BibitemOpen
  \bibfield  {author} {\bibinfo {author} {\bibfnamefont {S.~R.}\ \bibnamefont
  {{Hinton}}},\ }\bibfield  {title} {\bibinfo {title} {{ChainConsumer}},\
  }\href@noop {} {\bibfield  {journal} {\bibinfo  {journal} {The Journal of
  Open Source Software}\ }\textbf {\bibinfo {volume} {1}},\ \bibinfo {eid}
  {00045} (\bibinfo {year} {2016})},\ \bibinfo {note} {(See also
  \url{https://samreay.github.io/ChainConsumer/})}\BibitemShut {NoStop}%
\bibitem [{\citenamefont {Eadie}\ \emph {et~al.}(1971)\citenamefont {Eadie},
  \citenamefont {Drijard}, \citenamefont {James}, \citenamefont {Roos},\ and\
  \citenamefont {Sadoulet}}]{pull_reference}%
  \BibitemOpen
  \bibfield  {author} {\bibinfo {author} {\bibfnamefont {W.~T.}\ \bibnamefont
  {Eadie}}, \bibinfo {author} {\bibfnamefont {D.}~\bibnamefont {Drijard}},
  \bibinfo {author} {\bibfnamefont {F.}~\bibnamefont {James}}, \bibinfo
  {author} {\bibfnamefont {M.}~\bibnamefont {Roos}},\ and\ \bibinfo {author}
  {\bibfnamefont {B.}~\bibnamefont {Sadoulet}},\ }\href@noop {} {\emph
  {\bibinfo {title} {Statistical Methods in Experimental Physics}}}\ (\bibinfo
  {publisher} {North Holland},\ \bibinfo {year} {1971})\ pp.\ \bibinfo {pages}
  {277--278}\BibitemShut {NoStop}%
\bibitem [{\citenamefont {Demortier}\ and\ \citenamefont
  {Lyons}(2002)}]{pull_LucDemortier}%
  \BibitemOpen
  \bibfield  {author} {\bibinfo {author} {\bibfnamefont {L.}~\bibnamefont
  {Demortier}}\ and\ \bibinfo {author} {\bibfnamefont {L.}~\bibnamefont
  {Lyons}},\ }\href@noop {} {\emph {\bibinfo {title} {Everything you always
  wanted to know about pulls}}}\ (\bibinfo {year} {2002})\ \bibinfo {note}
  {(CDF/ANAL/PUBLIC/-5776)}\BibitemShut {NoStop}%
\bibitem [{\citenamefont {Workman}\ and\ \citenamefont {Others}(2022)}]{PDG}%
  \BibitemOpen
  \bibfield  {author} {\bibinfo {author} {\bibfnamefont {R.~L.}\ \bibnamefont
  {Workman}}\ and\ \bibinfo {author} {\bibnamefont {Others}} (\bibinfo
  {collaboration} {Particle Data Group}),\ }\bibfield  {title} {\bibinfo
  {title} {{Review of Particle Physics}},\ }\href
  {https://doi.org/10.1093/ptep/ptac097} {\bibfield  {journal} {\bibinfo
  {journal} {PTEP}\ }\textbf {\bibinfo {volume} {2022}},\ \bibinfo {pages}
  {083C01} (\bibinfo {year} {2022})}\BibitemShut {NoStop}%
\bibitem [{\citenamefont {Abbott}\ \emph {et~al.}(2022)\citenamefont {Abbott}
  \emph {et~al.}}]{DES_Y3_3x2pt}%
  \BibitemOpen
  \bibfield  {author} {\bibinfo {author} {\bibfnamefont {T.~M.~C.}\
  \bibnamefont {Abbott}} \emph {et~al.} (\bibinfo {collaboration} {DES}),\
  }\href {https://doi.org/10.1103/PhysRevD.105.023520} {\bibfield  {journal}
  {\bibinfo  {journal} {Phys. Rev. D}\ }\textbf {\bibinfo {volume} {105}},\
  \bibinfo {pages} {023520} (\bibinfo {year} {2022})}\BibitemShut {NoStop}%
\bibitem [{\citenamefont {Hamana}\ \emph {et~al.}(2020)\citenamefont {Hamana}
  \emph {et~al.}}]{HSC_2020}%
  \BibitemOpen
  \bibfield  {author} {\bibinfo {author} {\bibfnamefont {T.}~\bibnamefont
  {Hamana}} \emph {et~al.} (\bibinfo {collaboration} {HSC}),\ }\href@noop {}
  {\bibfield  {journal} {\bibinfo  {journal} {PASJ}\ }\textbf {\bibinfo
  {volume} {72}},\ \bibinfo {pages} {16} (\bibinfo {year} {2020})}\BibitemShut
  {NoStop}%
\bibitem [{\citenamefont {Abbott}\ and\ \citenamefont
  {et~al.}(2023)}]{KiDS-DES}%
  \BibitemOpen
  \bibfield  {author} {\bibinfo {author} {\bibfnamefont {T.~M.~C.}\
  \bibnamefont {Abbott}}\ and\ \bibinfo {author} {\bibfnamefont {M.~A.}\
  \bibnamefont {et~al.}} (\bibinfo {collaboration} {Dark Energy Survey and
  Kilo-Degree Survey Collaboration}),\ }\bibfield  {title} {\bibinfo {title}
  {{DES Y3 + KiDS-1000: Consistent cosmology combining cosmic shear surveys}},\
  }\href@noop {} {\  (\bibinfo {year} {2023})},\ \Eprint
  {https://arxiv.org/abs/2305.17173} {arXiv:2305.17173 [astro-ph.CO]}
  \BibitemShut {NoStop}%
\bibitem [{\citenamefont {Holm}\ and\ \citenamefont
  {et~al}(2023)}]{eBoss_study}%
  \BibitemOpen
  \bibfield  {author} {\bibinfo {author} {\bibfnamefont {E.~B.}\ \bibnamefont
  {Holm}}\ and\ \bibinfo {author} {\bibfnamefont {L.~H.}\ \bibnamefont
  {et~al}},\ }\href {https://doi.org/10.1103/PhysRevD.108.123514} {\bibfield
  {journal} {\bibinfo  {journal} {Phys. Rev. D}\ }\textbf {\bibinfo {volume}
  {108}},\ \bibinfo {pages} {123514} (\bibinfo {year} {2023})}\BibitemShut
  {NoStop}%
\end{thebibliography}%

\end{document}